\documentclass[a4paper,fleqn]{cas-sc}

\usepackage{times}
\usepackage[normalem]{ulem}
\usepackage{amsmath}
\usepackage{verbatim}
\usepackage{hyperref}
\usepackage{setspace}
\usepackage{lineno}
\usepackage{graphicx}
\usepackage{color}
\usepackage{lscape}
\usepackage{wasysym}
\usepackage[authoryear]{natbib}
 \usepackage{setspace} 
\graphicspath{{Figures/}}
\begin{document}
\newcommand{\CC}{\xi}
\newcommand{\CCp}{\hat{\CC}}
\newcommand{\FF}{{\Xi}}
\shorttitle{Evolution of the Pluto-Charon Binary}
\shortauthors{A.Bagheri et~al.}
\title[mode = title]{The Tidal-Thermal Evolution of the Pluto-Charon System}
\author{Amirhossein Bagheri$^1$}
\author{Amir Khan$^{1,2}$}
\author{Frederic Deschamps$^3$}
\author{Henri Samuel$^4$}
\author{Mikhail Kruglyakov$^{5,6}$}
\author{Domenico Giardini$^1$}
\address{$^1$Institute of Geophysics, ETH Zürich, Zürich, Switzerland}
\address{$^2$Physik-Institut, University of Zürich, Zürich, Switzerland}
\address{$^3$ Institute of Earth Sciences, Academia Sinica, 128 Academia Road, Sector 2, Nangang, Taipei 11529, Taiwan}
\address{$^4$Institut de Physique du Globe de Paris, CNRS, Université de Paris, Paris, France}
\address{$^5$University of Otago, Dunedin, New Zealand}
\address{$^6$Geoelectromagnetic Research Center, Institute of Physics of the Earth, Moscow}

\begin{abstract}
The existence of subsurface oceans on the satellites of the giant planets and Trans-Neptunian objects has been predicted for some time. Liquid oceans on icy worlds, if present, exert a considerable influence on the dynamics of the ice-ocean system and, because of the astrobiological potential, represent an important objective for future missions to the outer solar system. The Pluto-Charon system is representative of an icy moon orbiting a dwarf planet that is believed to have formed from the remnants of a giant impact. The evolution of icy moons is primarily controlled by the mode and efficiency of heat transfer through the outer ice shell, which is influenced by the presence of impurities, by tidal dissipation in the ice shell, and by the radioactive element budget in the silicate core. 
Previous studies on the evolution of the Pluto-Charon system generally considered either only the thermal or the tidal evolution, and in the cases where both were considered, the important effect of the presence of impurities in the liquid oceans was not addressed. Here, we consider the joint tidal-thermal evolution of the Pluto-Charon system by combining a comprehensive tidal model that incorporates a viscoelastic description of the tidal response with a parameterized thermal convection model developed for icy worlds. This approach enables an extensive analysis of the conditions required for the formation and maintenance of subsurface liquid oceans up to the present. Our results show that because of relatively fast circularization and synchronization of the orbits of Pluto and Charon, tidal heating is only important during the early stages of evolution ($<$1~Myr). As part of our study, we test the sensitivity of our results to a number of parameters, including initial orbital and thermal parameters. In all the studied cases, oceans on Pluto are always predicted to remain liquid to the present, ranging in thickness from 40~km to 150-km, whereas oceans on Charon, while in-place for close to 4~Gyr, have solidified. This is supported by New Horizons observations of primarily extensional faults on Pluto and both extensional and compressional faults on Charon. 
\end{abstract}

\maketitle
\doublespacing
\section{Introduction}
The Kuiper belt harbours numerous planetary objects of diverse internal structure and surface features, including the dwarf planet Pluto. Pluto hosts five known satellites named Charon, Kerberos, Hydra, Nix, and Styx, of which Charon is by far the largest. The Pluto system is, because of Pluto's size and relative brightness, presently the best-studied of all of the Trans-Neptunian objects (TNOs) \citep{Hussmann_etal06}. This is a consequence of a protracted history of Earth-based remote sensing \citep{MalhotraWilliams97,Dobrovolskis_etal97,olkin_etal03} and not least the flyby of the New Horizons spacecraft in 2015 \citep{spencer_etal20}. Pluto was found to display both a complex and an active geology that encompasses an extensive range of surface ages and a dynamic linkage between surface and atmosphere \citep{Nimmo_etal17,spencer_etal20}. In comparison, Charon appears to be both geologically and compositionally distinct to Pluto \citep{spencer_etal21}. The Pluto-Charon mass ratio was also observed to be much higher than that of the Earth-Moon system, which is untypical of Planet-satellite systems in our Solar System.
Charon, in analogy with the Earth-Moon system, is believed to have formed as a result of a collision between Pluto and a Kuiper belt object \citep{Canup11, Sekine_etal17, arakawa_etal19}. Like Earth's Moon, Charon would initially have been closer to Pluto, but because of tidal dissipation within the two bodies, Charon would have been driven further away until it reached its current synchronous state \citep{Dobrovolskis_etal97}.

Generally speaking, tidal evolution drives planetary systems towards equilibrium states by damping their orbital eccentricity, and forcing the spin rates towards stable synchronous rotation, while adjusting the separation between the planetary objects. 
During the deceleration of the spin of the planetary objects, heat is produced by friction, which, in addition to radiogenic heating, changes the thermal structure of the planet. As the planetary objects thermally evolve, their interior properties change, which in turn, influences their tidal response \citep[e.g.,][]{Robuchon_Nimmo11,Saxena_etal18, Samuel_etal19, Bagheri_etal21, Renaud_etal21}. Hence, the tidal and thermal evolution co-modulate, necessitating their joint consideration in evaluating the evolution of planetary systems. Because the amount of tidal dissipation depends on the thermal state, physical structure, and orbital parameters of the objects (distance, eccentricity, and spin and orbit rates), these all need to be studied within a single framework.  

In the context of tidal dissipation and the associated generation of heat, the New Horizons mission found indications that Pluto and Charon may harbour subsurface oceans beneath their ice-covered surfaces \citep{Nimmo_etal17,olkin_etal17}. The possibility that both bodies are able to sustain a liquid ocean beneath an icy cover further makes them prime targets for the search for extra-terrestrial life \citep{vance_etal18}.
For an ocean to form and remain liquid, the presence of long-lived heat sources is  required, of which radioactive and tidal heating are the most significant contributors \citep{McKinnon_etal97,HussmannSpohn04,Schubert_etal10,Robuchon_Nimmo11,Saxena_etal18}. Consequently, understanding the long-term thermal and tidal evolution of planetary systems is a central tenet in evaluating the possibility for the existence of a present-day subsurface ocean and, in turn, the astrobiological potential of the outer Solar System \citep{Mckinnon06,Mottl_etal07,vance_etal07}.


Several studies have addressed the evolution of the Pluto-Charon system in the context of coupled thermal-orbital evolution models \citep[e.g.,][]{Robuchon_Nimmo11,barr_colins15,hammond_etal16,desch_Neveu17,Saxena_etal18}. The role of tidal heating in the evolution of Kuiper belt Objects, including Pluto-Charon, was found to be comparable to and even higher than the heat produced by the radioactive decay of long-lived isotopes. \cite{Saxena_etal18}, for example, observed that subsurface oceans containing a small amount of impurities and tidal heating due to initially high spin rates may enable liquid water and cryovolcanism to persist until the present \citep{moore_etal16,Neveu_etal15,beyer_etal19}. Yet, these studies generally did not consider dissipation in both bodies and relied on tidal evolution models that are inadequate for the case of a highly eccentric and non-synchronously rotating system \citep{Bagheri_etal21,Renaud_etal21}. Specifically, tidal models that truncate eccentricity functions to $e^2$ (where $e$ is eccentricity) on eccentric orbits ($>$0.1) impart changes in spin rate evolution, spin-orbit resonances, and errors in heating rates that typically increase significantly for very high eccentricity ($>$0.5), which is not observed when higher-order terms are included. Moreover, while the aforementioned studies 
found the evolution to be fast ($\leq$1~Myr), incorporation of non-synchronous rotation can slow the evolution down if higher-order spin-orbit resonances are encountered \citep[e.g.,][]{Saxena_etal18}.

With this in mind, it is the purpose here to build upon and extend earlier studies on the evolution of the Pluto-Charon system by combining the comprehensive tidal model of \citet{Bagheri_etal21} that incorporates a proper viscoelastic description of the tidal response of icy worlds with the parameterized thermal convection models of \citet{Deschamps_Vilella21} (icy crust) that takes into account the effect of impurities in the liquid ocean and \citet{Samuel_etal19} (silicate core). This approach will enable an extensive analysis of 1) the tidal-thermal evolution of the Pluto-Charon system, allowing us to assess the relative role of radiogenic to tidal heating and 2) the conditions required for the formation and maintenance of subsurface liquid oceans up until the present. More generally, the methodology and results presented here can be exploited to understand the evolution of icy satellites and easily be extended to TNOs and exoplanets.

The manuscript is arranged as follows. In section 2, we describe current observations and the constraints they provide on the interior structure of Pluto and Charon; in section 3, we detail the thermal evolution and tidal models; in section 4, we present and discuss the results. 

\section{Pluto and Charon}\label{interior_properties}
\subsection{Interior}
Direct information bearing on the interior structure of Pluto and Charon is, as noted, scarce. Based on New Horizons observations, mass and radius of the two objects could be determined, allowing for the estimation of their mean densities \citep{Nimmo_etal17} (see Table~\ref{tab:orbital_params}). The bulk densities of Pluto and Charon are clearly higher than that of ice and therefore the interiors of the two bodies must be composed of denser material such as silicates and possibly also metals \citep{mckinnon_etal08}. The mean density of Pluto indicates a rock fraction of about 2/3, the rest possibly consisting of ice \citep{mckinnon_etal17}, whereas the lower density of Charon suggests a slightly lower rock/mass fraction. While porosity affects density, it is not expected to be able to account for the density difference between the two bodies \citep{mckinnon_etal17,Bierson_etal18}.

Pluto is believed to have formed from the hydrated silicate cores and icy material of the mantles of two impacting objects, both of which were already differentiated \citep{Desch15} or partially differentiated \citep{Canup11,desch_Neveu17}. Despite the lack of measurements of the moments of inertia, New Horizons observations indicate that both Pluto and Charon are most probably differentiated \citep{Stern15,stern_etal18,spencer_etal21}. The observations include surface spectra dominated by ices and lack of compression in the surface geological record that would otherwise be expected, were the interiors undifferentiated as a consequence of the formation of deeper and denser high-pressure ice as Pluto and Charon cooled
\citep{mckinnon_etal17, Grundy_etal16,hammond_etal16}. Instead, the observations indicate the presence of extensional tectonic features associated with
the expansion that occurred when an early ocean froze above a differentiated interior \citep{Stern15, moore_etal16, Beyer_etal17}.

On account of a substantial rock fraction combined with a relatively large size, it has been estimated that radioactive heat production within Pluto would be sufficient to melt a conductive ice shell and maintain a global subsurface ocean until the present day \citep{hammond_etal16,Bierson_etal18}. In contrast, a convective ice shell would allow for rapid removal of heat that an ocean would never develop \citep{Robuchon_Nimmo11}. Thus, the presence of an ocean today would suggest a cold, rigid, and conductive ice shell that allows for little interaction between ocean and surface. By the same argument, Charon is not expected to host a subsurface ocean today, although the latter might have developed at some point earlier in its evolution \citep{desch_Neveu17,Bierson_etal18}. In line herewith,
imaged extensional tectonic features are evidence of strains \citep{Beyer_etal17} that are expected if an ocean refreezes \citep{spencer_etal20}.
If Charon's initial orbit was non-circular, it would also have experienced an early episode of tidal heating and stress generation \citep{Rhoden_etal15}. 

\begin{table}
\begin{center}
\begin{tabular}{ll|cc}
\hline
\textbf{Parameter}        & \textbf{Symbol}                     & \textbf{Pluto}       & \textbf{Charon}       \\ \hline
\multicolumn{1}{l}{Radius (m)}  &  $R$  & 1.1883$\times$ 10$^{6}$      & 0.606$\times$ 10$^{6}$         \\
\multicolumn{1}{l}{Mass (kg)}    & $M$  & 1.328$\times$ 10$^{22}$ & 1.603$\times$ 10$^{21}$ \\
\multicolumn{1}{l}{Density (kg/m$^3$)} & $\rho$ & 1854 & 1701 \\
\multicolumn{1}{l}{Semi-Major Axis (m)}& $a$ & --                  & 19.596$\times$ 10$^{6}$       \\
\multicolumn{1}{l}{Spin period~(day)} & $\theta$ &  6.387  & 6.387       \\ 
\multicolumn{1}{l}{Eccentricity}& $e$ & --                  & $\sim$0       \\ 
\multicolumn{1}{l}{Inclination}& $i$ & --                  & $\sim$0       \\ 
\hline \hline
\end{tabular}
\caption{Properties of the Pluto-Charon system \citep{Nimmo_etal17, Stern15}}
\label{tab:orbital_params}
\end{center}
\end{table}

\begin{figure}  
\begin{center}
 \includegraphics[width=1.\textwidth]{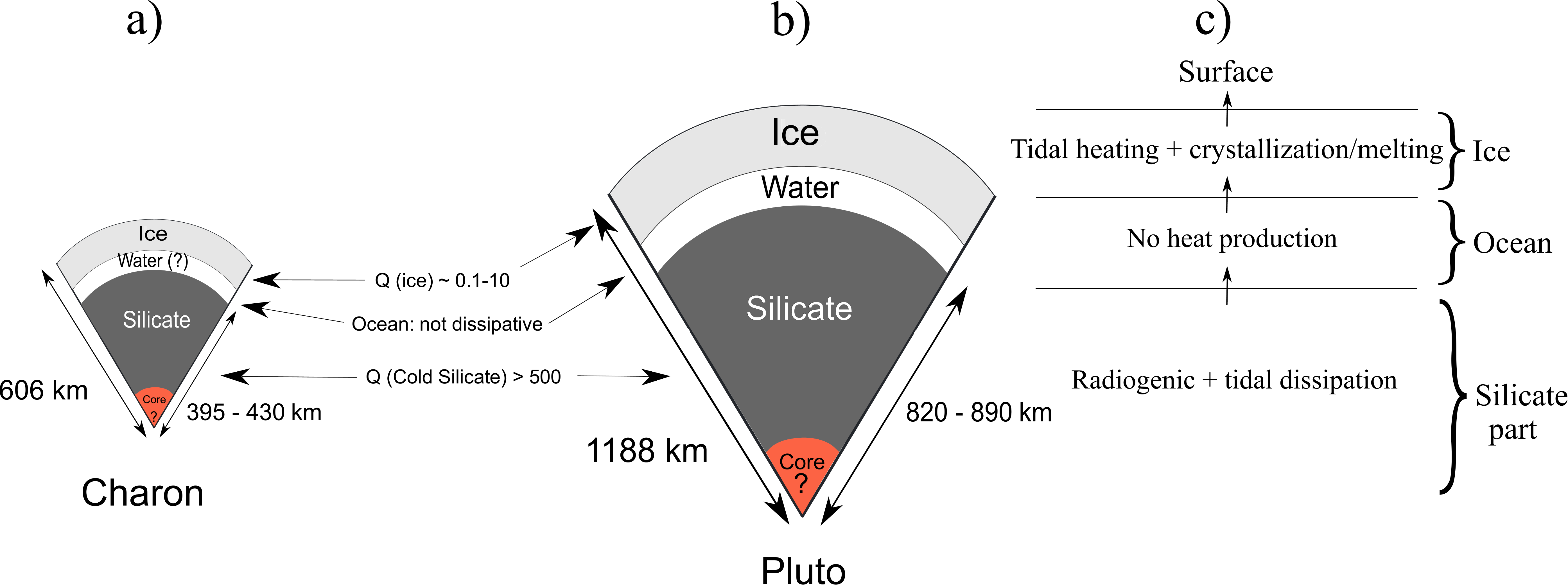}
\end{center}
\caption{\label{fig:density} Cross-sections of plausible interior structure models for Charon (a) and Pluto (b). In (a) and (b), the thicknesses of the silicate parts (and therefore also of the ice layers) range from 395--430~km for Charon and 820--890~km for Pluto, respectively, to keep the mean densities (Table~\ref{tab:orbital_params}) of the bodies fixed. Attenuation values (Q) are also indicated for the various layers. (c) illustrates the location of the main heat sources that are considered in this study, while arrows indicate the direction of the heat flow.}
\end{figure}

Plausible present-day interior structure models for Pluto and Charon are shown as cross-sections in Figure~\ref{fig:density} \citep{Stern15,Bierson_etal18,mckinnon_etal17,Nimmo_etal17,Rhoden20}. 
While a small iron core is indicated in the cross-section, we consider the core to be part of the silicate part hereafter, because of the negligible effect of a tiny core on the evolution. Given the dearth of direct observations on the interior, the radial extent of each layer is not well-constrained, which is reflected in the thickness ranges indicated in the cross-sections.
Estimations of the density and volume of the silicate part and ice/water layer are such that the measured mean densities for Pluto and Charon (Table~\ref{tab:orbital_params}) are satisfied. 
For the models of Pluto and Charon adopted here, we vary the density of the silicate core between 3000--3600~kg/m$^3$, as consequence of which the radius of the silicate part
varies in the range of 820--890~km for Pluto and 395--430~km for Charon, to satisfy the mean density of the bodies. In turn, this implies an ice/water thicknesses of 295--370~km and 175--210~km for Pluto and Charon, respectively. 
Physical properties for each layer are compiled in Table~\ref{tab:interior_props}. 

\begin{table}
\begin{center}
\begin{tabular}{lc|c}
\hline
\textbf{Parameter/unit}                             & \textbf{Symbol}       & \textbf{Value}   \\ \hline
\multicolumn{1}{l}{Ice}      &       &           \\
\multicolumn{1}{l}{\quad Shear modulus ~(GPa)   } & $\mu_{ice}$                  & 4.8      \\ 
\multicolumn{1}{l}{\quad Bulk modulus ~(GPa)} & $\kappa_{ice}$                  & 10.5     \\ 
\multicolumn{1}{l}{ \quad  Density~(kg/$\rm m^3$)}      & $\rho_{ice}$ & 920  \\
\multicolumn{1}{l}{Water } &                   &        \\
\multicolumn{1}{l}{\quad Shear modulus ~(GPa)   } &  $\mu_{w}$              & 0       \\ 
\multicolumn{1}{l}{\quad Bulk modulus ~(GPa)} &  $\kappa_{w}$                  & 2.2     \\ 
\multicolumn{1}{l}{\quad Density~( kg/$\rm m^3$) } &  $\rho_{w}$            & 1000       \\
\multicolumn{1}{l}{Silicate } &        &        \\
\multicolumn{1}{l}{\quad Shear modulus  ~(GPa)  } &  $\mu_{sil}$                  & 60      \\ 
\multicolumn{1}{l}{\quad Bulk modulus ~(GPa)} &  $\kappa_{sil}$                  & 120     \\ 
\multicolumn{1}{l}{\quad Density ~(kg/$\rm m^3$)} & $\rho_{sil}$                  &  3000--3600   \\
\hline \hline
\end{tabular}
\caption{Properties of different layers for the reference model}
\label{tab:interior_props}
\end{center}
\end{table}

\subsection{Orbital properties}
The observed orbital properties, including semi-major axis ($a$), eccentricity ($e$), inclination ($i$), and spin period ($\theta$) of Pluto and Charon are given in Table~\ref{tab:orbital_params}. The two bodies are presently tidally locked to each other, i.e., their spin periods are equal to their orbital period around their center of mass, ensuring that each always presents the same face to the other, and their respective orbits about the center-of-mass are almost perfectly circular. The present-day eccentricity and inclination of the orbit is very close to zero \citep{buie_etal12,Stern15,Nimmo_etal17,Brozovic15}. This implies that there are no significant active tidal forces acting on the two bodies, and hence, negligible tidal dissipation occurring within either object at the present.

\section{Methods}
\subsection{Thermal evolution of Pluto and Charon}\label{Thermal_evolution}

The thermal evolution of a planetary object is controlled by the processes that produce heat and those responsible for transferring the heat within the different planetary envelopes and to the surface, either through radiation, conduction or convection. The main heat sources in a binary system are:
\begin{enumerate}
    \item the impact heating associated with accretion during planet formation,
    \item the gravitational energy released during planetary differentiation, 
    \item radiogenic heating in the silicate component from the decay of long-lived radioactive isotopes (U, Th, and K), and 
    \item tidal heating due to viscoelastic dissipation.
\end{enumerate}
Of these sources, only 3) and 4) are of relevance for the long-term evolution of the planet. Sources related to 1) and 2) are principally linked to the very earliest stages of planetary accretion. 
A large part of the accretional heat is likely to have radiated immediately into space \citep{hussmann_etal10}, whereas the energy release during early differentiation is estimated to be 10\% at most of the accretional energy \citep{Schubert_etal86} and consequently, negligible in comparison to the contribution from e.g., radiogenic heating \citep{HussmannSpohn04}. Since we consider the evolution of the Pluto-Charon system from the time postdating formation and differentiation, we neglect their contributions in the following. In summary, the main heat sources affecting the thermal evolution of Pluto and Charon are those associated with radiogenic heating in the silicate parts and tidal dissipation in the ice shell. Tidal dissipation in the solid core is negligible because of the low temperatures (see Section~\ref{sec:viscoelast}).

As illustrated in Figure~\ref{fig:density}c, the evolution of the ice shell is computed based on the energy balance between 1) the heat produced in the silicate part (radiogenics) and entering the ice shell from below; 2) the heat produced in the shell (tides and crystallization/melting of ice); and 3) the heat that can be transported from within the ice shell to the surface. Any ocean that is either initially present or forms during the course of the evolution, is assumed to be adiabatic. 

The properties of the outer ice shells of Pluto and Charon that are made up of ice Ih may allow thermal convection to operate within them \citep{Hussmann2007}. Details of this mechanism depend on the physical properties of the system. The rheology of ice plays an important role in heat transfer through the outer ice shell and depends strongly on the material temperature. Because the temperature-dependence of the viscosity of ice Ih is large \citep[e.g.,][]{Durham2010}, convection in the outer ice shell may occur in the so-called stagnant-lid regime, for which the amount of heat that can be transported to the surface is reduced because of the presence of a thermally conductive and rigid (high viscosity) lid at the top of the system \citep[e.g.,][]{MoresiSolomatov1995,DavailleJaupart1993}. Stagnant-lid convection strongly influences the heat flux and interior temperature, which both depend on the top-to-bottom  thermal viscosity contrast of the ice layer. The release of heat through tidal dissipation (to be discussed in Section~\ref{Tidal_evolution}) within the ice shell further influences the properties of the ice shell and its ability to transport heat to the surface, as suggested by numerical simulations of mixed-heated (basal and internal) thermal convection \citep[e.g.,][]{TravisOlson1994,Deschamps2010}. 

In the ice and water layers we model thermal evolution using the parameterized convection model of \citet{GrassetSotin96}. This approach uses scaling laws derived from simulations of thermal convection to estimate the average temperature within the ice layer and the heat flux at its bottom. The growth of the ice shell (thickness) is then estimated from the difference between the heat flux coming from the silicate part (hereinafter core) and the heat flux entering the ice shell. At the boundary between the ice shell and the subsurface ocean, energy conservation is written  
\begin{equation}\label{eq:heat}
\frac{dr_{bot}}{dt}\bigg[\rho_w C_w \bigg(  \frac{\partial T_{bot}}{\partial r}-\frac{\partial T_{ad}}{\partial r}   \bigg) \frac{r_{bot}^3 - r_c^3}{3} - \rho_I L_I r_{bot}^2 \bigg] = r_{bot}^2 \phi_{bot} - r_c^2 \phi_c,   
\end{equation}
where $t$ is time, $T_{bot}$ is the temperature at the bottom of the ice layer given by the liquidus of the ocean, $\phi_{bot}$ is the heat flux at the bottom of the ice layer, $r_{bot}$ is the radius of the bottom of the ice layer, $r_c$ is the core radius, $\phi_c$ is the heat flux at the top of the core, $\rho_w$ and $C_w$ are the liquid water density and heat capacity, respectively, and $\rho_I$ and $L_I$ the density and latent heat of fusion of ice Ih, respectively. $T_{ad}$ is the adiabatic temperature in the ocean and is given by
\begin{equation}
    T_{ad} (r) = T_{bot}(r_{bot}) \bigg[ 1-\frac{\alpha_w}{\rho_w C_w} \rho_I g (r-r_{bot})\bigg],
\end{equation}
where $\alpha_w$ is the coefficient of thermal expansion of liquid water, $g$ is gravity, $\rho_I$ is the density of ice Ih, and $r_{top}$ is the radius of the top of the ocean layer. We model the ocean as an inviscid fluid layer through which heat is transported immediately from the bottom to the top of the ocean. We assume a uniform temperature profile for the ocean and ignore the tiny increase in temperature resulting from the adiabatic temperature gradient. 
Solving Eq.~\ref{eq:heat} for $r_{bot}$, i.e., ice layer thickness, requires knowledge of the heat fluxes at the top of the core ($\phi_{c}$ ) and at the bottom of the ice shell ($\phi_{bot}$) with time. 

The heat flux at the top of the core is obtained by modeling its thermal evolution, which is governed by solving the time-dependent heat diffusion equation following the approach of \citet{Samuel_etal19}. We assume that the core has a carbonaceous chondrite composition \citep{lodders03} and 
consider four radioactive elements: $\rm ^{235}U$, $\rm ^{238}U$, $\rm ^{232}$Th, and $\rm ^{40}$K. The averaged radial temperature profile of the core ($T_{c}$) and the heat flux at its top ($\phi_c$) are calculated by solving the heat diffusion equation 
using
\begin{equation}\label{thermal}
    \rho_c C_c \frac{\partial T_{c}}{\partial t} = \frac{1}{r^2}\frac{\partial}{\partial r} \bigg(r^2 k_c \frac{\partial T_{c}}{\partial r}\bigg) + H(t),
\end{equation}
where $\rho_c$, $C_c$, and $k_c$ are the density, the specific heat, and thermal conductivity of the core, respectively, and the volumetric internal heating rate, $H$ is the heat caused by tidal dissipation. The intial conditions are the initial thermal profiles for either body (shown in Figure~\ref{fig:temps}). Values for all parameters are listed in Table~\ref{tab:thermal_parameters}.

\begin{table} 
\begin{center}
\begin{tabular}{lc|c}
\hline
\textbf{Parameter (Unit)}           &    \textbf{Symbol}          & \textbf{Value/expression}      \\ \hline
\multicolumn{1}{l}{\textit{Ice Ih properties}}  & & \\
\multicolumn{1}{l}{ \quad Surface temperature $\rm (K)$}   & $T_{surf}$ & 40 \\
\multicolumn{1}{l}{ \quad Thermal expansion $\rm( 1/K)$}   & $\alpha_I$ & 1.56$\times$ 10$^{-4}$ \\
\multicolumn{1}{l}{ \quad Thermal conductivity $\rm (W/m/K)$} & $k$   & 2.6  \\ 
\multicolumn{1}{l}{ \quad Heat capacity $\rm (J/kg)$} & $C_p$ & 7.037$T$ + 185           \\ 
\multicolumn{1}{l}{ \quad Thermal diffusivity $\rm (m^2/s)$}       &  $\kappa_I$ & $k/{(\rho_I C_p)}$       \\ 
\multicolumn{1}{l}{ \quad Latent heat of fusion $\rm (kJ/kg)$}      & $L_I$ & 284       \\ 
\multicolumn{1}{l}{ \quad Reference bulk viscosity $\rm (Pa. s)$}    & $\eta_{ref}$ & 10$^{14}$       \\ 
\multicolumn{1}{l}{ \quad Activation energy $\rm (kJ/mol)$}           & $E^*$  & 60       \\ 
\multicolumn{1}{l}{ \quad Initial Thickness (Pluto) $\rm (km)$}    & $D_{init}^{ice}$ & 298       \\ 
\multicolumn{1}{l}{ \quad Initial Thickness (Charon) $\rm (km)$}    & $D_{init}^{ice}$ & 171       \\ 
\multicolumn{1}{l}{\textit{Liquid water properties}}  & &      \\ 
\multicolumn{1}{l}{ \quad Initial ammonia content $\rm (wt\%)$}    & $A_{init}$ & 3       \\ 
\multicolumn{1}{l}{ \quad Thermal expansion $\rm (1/K)$}               &  $\alpha_w$ & 3$\times$10$^{-4}$       \\ 
\multicolumn{1}{l}{ \quad Heat capacity $\rm (J/kg)$}              &$C_w$  & 4180       \\ 
\multicolumn{1}{l}{ \quad Initial Thickness (Pluto) $\rm (km)$}    & $D_{init}^{w}$ & 15       \\ 
\multicolumn{1}{l}{ \quad Initial Thickness (Charon) $\rm (km)$}    & $D_{init}^{w}$ & 15       \\ 
\multicolumn{1}{l}{\textit{Silicate core properties}}   & & \\
\multicolumn{1}{l}{ \quad Thermal conductivity $\rm (W/m/K)$}        &      $k_c$     & 2.4       \\
\multicolumn{1}{l}{ \quad Heat capacity $\rm (J/kg/K)$}        &      $C_c$    & 1100       \\
\multicolumn{1}{l}{ \quad Core radius (Pluto) $(\rm km)$}        &      $R_c$    & 875       \\
\multicolumn{1}{l}{ \quad Core radius (Charon) $(\rm km)$}        &      $R_c$    & 420       \\
\multicolumn{1}{l}{ \quad Density $(\rm kg/m^3)$}        &      $\rho_c$    & 3100       \\
\hline \hline
\end{tabular}
\caption{Thermal properties of different layers of the system}
\label{tab:thermal_parameters}
\end{center}
\end{table}

Depending on its properties, principally viscosity and thickness, the ice shell may transport heat either by conduction or by convection, leading to two different estimates of heat flux at its bottom. For a static, thermally conductive ice shell, the bottom heat flux is simply given by the the static heat equation in spherical geometry. If, by contrast, convection operates within the ice shell, the bottom heat flux is deduced from scaling laws derived from simulations of mix-heated stagnant-lid convection (\cite{Deschamps_Vilella21}). In practice, we compute both the conductive and convective heat fluxes, $\phi_{cond}$ and $\phi_{conv}$, and assume that convection operates if $\phi_{conv}$ > $\phi_{cond}$.

To relate the ice shell Rayleigh number, $Ra$, which measures the vigour of convection, and the viscous temperature scale $\gamma$, which controls the variation of viscosity with temperature, to the bottom heat flux, we use the following relations
\begin{equation}\label{eq:Rayleigh}
    Ra = \frac{\alpha_I \rho_I g \Delta T D^3}{\eta_{m} \kappa_I},
\end{equation}
\begin{equation}\label{eq:gamma}
    \gamma = \frac{E^* \Delta T}{R_{gas} {T_{m}}^2},
\end{equation}
\begin{equation}\label{eq:heatflux_conv}
    \phi_{bot} = 1.46 \frac{k_{I}\Delta T}{D f^2}  \frac{Ra^{0.27}}{\gamma^{1.21}},
\end{equation}
where $\alpha_{I}$, $\rho_{I}$, $E^*$, $\kappa_{I}$, and $k_{I}$ are the thermal expansion coefficient, density, activation energy, thermal diffusivity, and thermal conductivity of solid ice, respectively, $g$ is the gravitational acceleration of the body, $T_{m}$ is the interior temperature of the ice shell, $\eta_{m}$ the ice viscosity at this temperature, $D$ is the ice shell thickness, $\Delta T$ = $T_{bot}-T_{surf}$ the temperature jump across it, $f$ the ratio between its inner and outer radii ($r_{bot}$/$R$), and $R_{gas}$ is the ideal gas constant. Surface temperature is set to 40~K after \citet{McKinnon_etal97} and kept constant during the simulations. Numerical values for all parameters are listed in Table ~\ref{tab:thermal_parameters}. 

The interior temperature $T_{m}$ is defined as the averaged temperature within the well-mixed convective interior, i.e., excluding the stagnant lid and thermal boundary layers, and is given by the following relationship \citep[see][for details]{Deschamps_Vilella21},
\begin{equation}\label{eq:interior_temperature}
    T_{m} = T_{bot} - 1.23\frac{R_{gas} T_{m}^2}{E^* f^{1.5}} + (3.5 - 2.3f)\frac{1 + f + f^2}{3}\frac{\rho_{I} H D^2}{k_{I} \Delta T}\frac{\Delta T}{Ra^{0.25}},
\end{equation}
where $H$ is, as before, the internal heating due to tidal dissipation. Because Ra implicitly depends on $T_m$ (through the viscosity), Eq.~\ref{eq:interior_temperature} does not have an analytical solution. We solved Eq.~\ref{eq:interior_temperature} using a standard Newton-Raphson method. The interior viscosity is deduced from $T_{m}$ following
\begin{equation}\label{eq:interior_viscosity}
    \eta_{m} = \eta_{ref} \exp\bigg[\frac{E^*}{R_{gas} T_{ref}}\bigg(\frac{T_{ref}}{T_{m}}-1\bigg)\bigg],
\end{equation}
where $\eta_{ref}$ is the viscosity of pure water ice close to the melting point, which, in our case, is the temperature of the water liquidus at the bottom of the ice shell $T_{ref}$ = $T_{H2O,bot}$. The thermal conductivity of ice I, $k_{I}$, depends on temperature \citep[]{AndersonSuga1994}, but is here assumed constant throughout the ice shell. \citet{Deschamps21} showed that temperature-dependent conductivity has a strong impact on the thickness and thermal structure of the stagnant lid, but that the thermal evolution of the ice shell is nevertheless well-approximated by a homogeneous conductivity provided that the conductivity is considered at the temperature at the bottom of the ice shell, which, for most icy bodies spans the range 2.0--3.0~W/m/K. Here we assumed a value of 2.6~W/m/K \citep{Deschamps21}. 

We may also point out that the heat flux given by Eq.~\ref{eq:heatflux_conv} does not explicitly depend on $H$. It is, however, affected by $H$ through Ra and $\gamma$, which both depend on $T_{m}$ and thus on $H$. It is also worth noting that, while Eqs.~\ref{eq:heatflux_conv} and \ref{eq:interior_temperature} were built from simulations assuming a homogeneous distribution of $H$ throughout the ice shell, they also describe simulations with viscosity-dependent heating, provided that the maximum tidal dissipation (the highest $H$) occurs in the hottest regions \citep[]{Deschamps_Vilella21}.

In addition to the amount of internal heating within the ice shell, the presence of impurities within the subsurface ocean can affect the thermal evolution of theses bodies \citep[e.g.,][and references therein]{deschamps_sotin01,GrassetPargamin05}, by lowering the melting temperature of the ice shell and  thereby changing its crystallization behaviour. Associated herewith is an increase in bulk viscosity, which reduces the vigour of convection and the efficiency of heat transfer, further delaying the crystallization of the ice shell. New Horizons observations of the surfaces of Pluto and Charon reveal CH$_4$, N$_2$, and CO ices \citep{Grundy_etal16}. In addition, H$_2$O and NH$_3$ ices have also been detected that could have originated in the interior \cite{DalleOre_etal19}. Here, we consider ammonia ($\rm NH_3$) as the main impurity species  \citep[e.g.,][]{Desch09,waite_etal09,clark_etal14,Grundy_etal16,NimmoPappalardo16,dalle_etal18,beyer_etal19} and rely on the liquidus of the $\rm H_2O-NH_3$ system of \citet{deschamps_sotin01}. Other impurity compounds, in particular magnesium sulfates and methane, may also be present in icy world objects with qualitatively similar effects as ammonia \citep{vance_etal18,Kamata_etal19,vilella_etal20}. One thing to bear in mind in the context of the $\rm H_2O-NH_3$ system, is that only water crystallises until the eutectic composition is reached (32.2 vol\% NH$_3$). Impurities (NH$_3$) remain in the ocean, whereby their concentrations increase and, as a result, their impact on the ice shell as it thickens (by increasing viscosity).

Finally, to compute the thermal evolution of Pluto and Charon, we integrate Eq.~\ref{eq:heat} using an adaptive step-size-controlled 4th-order Runge-Kutta method, and we deduce the thermal properties of the ice shell (including its average temperature and the thickness of the stagnant lid) according to scaling laws developed in \citet{Deschamps_Vilella21}, to which the reader is referred for more details. As for the timing of the formation of the Pluto system, we assume 1) that the accretion process has been completed after the CAIs have formed, and therefore disregard the contribution from short-lived radionuclides, and 2) that Charon formed 200~Myr after the formation of Pluto \citep[e.g.,][]{Canup_etal21}. 
In connection with the initial thermal state of Pluto and Charon, we assume an initially homogeneous interior temperature, 
and consider both a cold- and a hot-start case \citep[e.g.,][]{Bierson_etal20,Renaud_etal21}. Cold-start (180~K) corresponds to an absent or very thin ocean layer (15~km thickness), whereas a hot-start commences right at the point where the ocean layer (280~km thickness) starts to crystallize, corresponding to $\sim$270~K for pure water and $\sim$250~K in the case contaminants are present. The tidal heating will be discussed in the next section.

\subsection{Tidal evolution of a highly eccentric non-synchronous rotating binary system} \label{Tidal_evolution}
\subsubsection{Tidal response}
An orbiting moon will raise a tide on a planet, resulting in an imposed potential $\Phi$ that causes the planet to deform. The deformation induces a gravitational potential $\psi$ given by
\begin{equation}\label{eq:love1}
\psi_l ({\bf{r}})= \left(\frac{R}{r}\right)^{l+1}
k_l \Phi_l({\bf{R}}, {\bf{r}}^*),
\end{equation}
where $R$ is the radius of the planet, ${\bf{R}}$ is a point on the planet's surface, ${\bf{r}}$ is a point exterior to ${\bf{R}}$, while ${\bf{r}}^*$ describes the position of the perturbing body, and
both $\Phi$ and $\psi$ are expanded in terms of spherical harmonics of degree $l$. The $k_\mathrm l$ are tidal Love numbers of degree $l$ and depend on the interior properties of the deformed body and determine the amplitude of its response \citep[e.g.,][]{Efroimsky12a}.

The expression for the potential of the tidally deformed planet (Eq.~\ref{eq:love1}) is valid for a purely elastic response of the planet. In this case, the tidal bulge
raised on the planet by the moon is aligned with the direction from the planet's centre towards the moon,  with no lagging. This ensures that the torques, applied by the moon on the planet and the opposite
torque with which the planet is acting on the moon, are zero. Consequently, in the elastic case, there is no influence of the planetary tides on the moon's orbital parameters (semi-major axis, eccentricity, and inclination), and, therefore, tidal heat is not generated in the planet.

Realistic objects, however, deviate from the purely elastic case, as a result of which the tidal bulge acquires a complex structure and is no longer aligned with the perturbing body. Following \citet{EfroimskyMakarov14}, we decompose the bulge over the tidal Fourier modes. This results in harmonics, of which some lag and some are in advance of the sub-satellite point, but each harmonic produces tidal heat. In this case Eq.~\ref{eq:love1} needs to be modified, in the time domain, to yield
\begin{equation}
\psi_l ({\bf{r}}, t) = \left(\frac{R}{r}\right)^{l+1}
\hat{\mathrm k}_l \Phi_l({\bf{R}}, {\bf{r}}^*),
\label{}
\end{equation}
where $\hat{\mathrm k}_l$ is the (linear) Love operator that maps the entire
history of the perturbation ($\Phi_l(t')$ over $t'\leq t$) on the value of $\psi$ at the present 
$t$. In the time domain, this can be written in the form of a convolution,
while in the frequency domain it is given as a product of Fourier components by
\begin{equation}
\bar{\psi}_l ({\bf{r}}, \omega^\mathrm{nm}_\mathrm{pq}) =
\left(\frac{R}{r}\right)^{l+1}
\bar{\mathrm k}_l(\omega^\mathrm{nm}_\mathrm{pq})
\bar{\Phi}_l ( {\bf{R}}, {\bf{r}}^*, \omega^\mathrm{nm}_\mathrm{pq}),
\label{eq:love2}
\end{equation}
where $\omega^\mathrm{nm}_\mathrm{pq}$ are the Fourier tidal modes (whose absolute values are
the physical forcing frequencies exerted in the material) and \{$^\mathrm{nm}_\mathrm{pq}$\} are integers
that are used to number the modes. 
In the latter expression, overbars are employed to denote complex Fourier components, i.e.,
\begin{equation}
\bar{\mathrm k}_l(\omega^\mathrm{nm}_\mathrm{pq}) =
\mathrm{Re}[\bar{\mathrm{k}}_l(\omega^\mathrm{nm}_\mathrm{pq})] +
i\,\mathrm{Im}[\bar{\mathrm{k}}_l(\omega^\mathrm{nm}_\mathrm{pq})] =
|\bar{\mathrm{k}}_l|~e^{-\epsilon_l(\omega^\mathrm{nm}_\mathrm{pq})}.
\label{eq:love3}
\end{equation}
In Eq.~\ref{eq:love2}, $\bar{\psi}_l ({\bf{r}}, \omega^\mathrm{nm}_\mathrm{pq})$ is lagging behind
$\bar{\Phi}_l ( {\bf{R}}, {\bf{r}}^*, \omega^\mathrm{nm}_\mathrm{pq})$ by the phase angle
$\epsilon_l(\omega^\mathrm{nm}_\mathrm{pq})$, which by convention is the negative argument
of the complex Love number $\bar{\mathrm k}_l(\omega^\mathrm{nm}_\mathrm{pq})$.
Replacing ${\bf{r}}$ by the position of the perturbing body ${\bf{r}}^*$, the
additional potential ``felt'' by the latter (Charon) is obtained.

As shown in \citet[e.g.,][]{EfroimskyMakarov14}, to each tidal mode $\omega^\mathrm{nm}_\mathrm{pq}$ corresponds an appropriate Fourier contribution that is proportional to the sine of the phase lag at that mode. By convention, the quantity inverse to the absolute value of this sine is termed the tidal
quality factor, $Q_l$, and defined as
\begin{equation}
\frac{1}{{Q_l}(\omega^\mathrm{nm}_\mathrm{pq})} =
\sin | \epsilon_l(\omega^\mathrm{nm}_\mathrm{pq}) | .
\label{dissip9}
\end{equation}

In the following, we concentrate on the tides related to degrees 2 and 3. While both $k_l$ and $Q_l$ depend on interior properties (density and rigidity), $Q_l$ is highly sensitive to viscosity and, therefore, to temperature (Eq.~\ref{eq:interior_viscosity}). We will briefly describe this in the next section. For the model specified in Table~\ref{tab:interior_props}, we obtain present-day degree-2 and -3 tidal Love numbers of $k_2=0.09$, $h_2=0.28$, and $l_2=0.058$ and $k_3=0.05$, $h_3=0.2$, and $l_3=0.06$, respectively. Equivalent Love numbers for Charon are $k_2=0.001$, $h_2 = 0.002$, and $l_2 = 0.001$ and $k_3=0.0006$, $h_3 = 0.0014$, and $l_3 = 0.0004$, respectively. The considerable difference between the Love numbers of the two objects is partly due to the larger size of Pluto and partly related to the presence of a large subsurface ocean within Pluto, which causes more ``flexibility" of Pluto relative to solid Charon.

\subsubsection{Viscoelastic dissipation}\label{sec:viscoelast}

Tidal heating takes place in the planetary bodies as a result of a variety of viscoelastic creep processes \citep[e.g.,][]{JacksonFaul10}.
Several rheological models (e.g., Maxwell, Andrade, Burgers, and Sundberg-Cooper) have been proposed to model the viscoelastic creep based on laboratory measurements  \citep[e.g.,][]{JacksonFaul10,SundbergCooper10} and have been employed to model dissipation in planetary bodies \citep[e.g.,][]{RobertsNimmo08, Harada_etal14, Efroimsky15, WilliamsBoggs15, MccarthyCooper16, RenaudHenning18, Khan_etal18, Bagheri_etal19, Tobie_etal19}. 
The Sundberg-Cooper viscoelastic model \citep{SundbergCooper10} is used in this study. This model is a composite viscoelastic model that includes features from both of the commonly-used Andrade and extended Burgers models, i.e., the ``response broadening" behavior and the experimentally observed secondary dissipation peak \citep{JacksonFaul10}. The Sundberg-Cooper model has been shown to cover the viscoelastic properties of both ice and silicate materials \citep{SundbergCooper10,caswell_etal15,caswel_cooperl16,MccarthyCooper16}. 
However, laboratory measurements of torsional forced oscillations of silicate materials show that below temperatures of about 600$\rm ^\circ C$, viscoelastic creep does not occur \citep{JacksonFaul10}. In the case of Pluto and Charon, temperatures of the silicate core do not exceed 500$\rm ^\circ C$ during the orbital evolution, even in a hot-start formation scenario \citep{Bierson_etal20}. Consequently, dissipation in the silicate part is negligible with quality factors generally exceeding 500. In contrast, ice is considerably more dissipative than silicate with quality factors as low as 0.1 \citep{MccarthyCooper16}. The Sundberg-Cooper viscoelastic model is described in more detail in Appendix~\ref{App:A}. Finally, tidal dissipation in the liquid ocean layer is assumed negligible. This will be further discussed in section~\ref{sec:discussion}. In summary, tidal dissipation in Pluto and Charon is predominantly taking place in their ice shells.

An important parameter of any viscoelastic model is the frequency exponent $\alpha$, which determines the variation of the tidal response with frequency. Higher $\alpha$ corresponds to larger dissipation as the period increases ($Q\sim \omega^{-\alpha}$), implying, for example, that as the two objects recede from one another, the stable synchronous state may be reached in a shorter time interval. The exact value of $\alpha$ is not well-constrained, but most laboratory measurements suggest values in the range 0.25-0.33, which is valid for silicate and icy materials \citep{JacksonFaul10,MccarthyCooper16}.
Here, we use $\alpha$$\approx$0.27,  
but we will also consider the effect of variations in $\alpha$ on the orbital evolution. 

\subsubsection{Tidal evolution model}\label{orbital_evolution}
After the putative Charon-forming giant impact \citep{Canup05,stern_etal06,arakawa_etal19}, the orbits of Pluto and Charon have been separated because of the transfer of angular momentum in analogy with the Earth-Moon system. The tidal evolution of Pluto and Charon consists of expansion of the post-impact orbit of the satellite around the dwarf planet, concomitantly with damping of the initially highly eccentric orbit and despinning of both objects from initially higher spin rates (Table~\ref{tab:orbital_params}).

Here, we employ the tidal evolution model of \citet{Bagheri_etal21}, which extends the Darwin-Kaula and \citet{BoueEfroimsky19} tidal models through the use of higher-order eccentricity functions and harmonic modes. As the initial orbit of Charon around Pluto is believed to have been highly eccentric \citep{Canup05}, the use of higher-order eccentricity functions and harmonic modes is essential. Here, we further extended our tidal model to include the case of non-synchronous rotation. Due to the spherical shape of both bodies, librational tides are not important.  

For a non-synchronous planet hosting a non-synchronous satellite without libration, the tidal rates of the semi-major axis $a$, eccentricity $e$, spin rate $\dot{\theta}$, inclination $i$, and tidally dissipated energy $E$ can be written in terms of the mean motion ($n$), planet and satellite masses ($M$ and $M^\prime$), the planet and satellite radii ($R$ and $R^\prime$), the quality functions ($K_l = k_l/Q_l$ and $K_l^{\,\prime}=k_l^{\,\prime}/Q_l^{\,\prime}$), and spin rates ($\dot{\theta}$, and $\dot{\theta}^{\,\prime}$):
\begin{equation}\label{eq1}
\frac{da}{dt}= \frac{G(M+M^{\prime})}{n~a^2}\frac{M^{\prime}}{M}\bigg( \frac{R}{a} \bigg)^5 \bigg[\underbrace{{\cal{F}} ( K_l,\,\dot\theta,\,n,\,e)}_{\text{tides on primary body}} + \underbrace{{\cal{F}} ( K^\prime_l,\,\dot\theta^{\prime},n,\,e)}_{\text{tides on satellite body}}\bigg],
\end{equation}
\begin{equation}\label{eq2}
\frac{de}{dt}=  \frac{G(M+M^{\prime})}{n~a^3}\frac{M^{\prime }}{M}\bigg( \frac{R}{a} \bigg)^5 \bigg[\underbrace{{\cal{L}} ( K_l,\,\dot \theta,\,n,\,e)}_{\text{tides on primary}} + \underbrace{{\cal{L}} ( K^\prime_l,\,\dot\theta^{\prime},n,\,e)}_{\text{tides on satellite}} \bigg]\;\;,
\end{equation}\
\begin{equation}\label{eq3} 
\bigg(\frac{d\dot{\theta}}{dt}\bigg)_{\text{primary}} = ~ \frac{G M^{\prime 2}}{a^3 M R^2}\bigg( \frac{R}{a} \bigg)^3 \bigg[\underbrace{ {\cal{G}} ( K_l,\,\dot \theta,\,n,\,e)}_{\text{tides on primary}}\bigg],\;\;
\end{equation}
\begin{equation}\label{eq4}
\bigg(\frac{di}{dt}\bigg)_{\text{primary}} =~n \sin i \frac{M^{\prime}}{M}\bigg( \frac{R}{a}\bigg)^5 \bigg[\underbrace{{{\cal{I}} (K_l,\dot{\theta},n,i,e)}}_{\text{tides on primary}}\bigg] ,\;\
\end{equation}
\begin{equation}\label{eq5}
\bigg(\frac{dE}{dt}\bigg)_{\text{primary}} = ~ \frac{GM^{\prime 2}}{a}\bigg( \frac{R}{a} \bigg)^5 \bigg[\underbrace{{\cal{H}} ( K_l,\,\dot \theta,\,n,\,e)}_{\text{tides on primary}} \bigg].\;\;
\end{equation}
Identical expressions for the time derivatives of $\dot \theta$, $i$, and $E$ of the secondary are obtained by replacing $R$, $K_l$, and $\dot \theta$ with those of the secondary, i.e., $R^{\prime}$, $K_l^{\,\prime}$, and $\dot \theta^{\prime}$, and interchanging $M$ and $M^{\prime}$. Detailed expressions for the functions ${\cal {F}}$, ${\cal {L}}$, ${\cal {I}}$, ${\cal {G}}$, and ${\cal {H}}$ are given in Appendix~\ref{App:B}. As apparent from Eqs.~\ref{eq1}--\ref{eq5}, our tidal model includes dissipation within both the primary (Pluto) and the satellite (Charon) contrary to previous studies that only accounted for dissipation in Pluto \citep{Robuchon_Nimmo11,barr_colins15,hammond_etal16,arakawa_etal19}. 

Our tidal model is also favored over the more commonly-used Constant Time Lag (CTL) and Constant Phase Lag (CPL) models \citep[e.g.,][]{Heller_etal11, cheng_etal14, Samuel_etal19, arakawa_etal19,Ferraz-Mello_20}. The CTL model erroneously implies that all the tidal strain modes experience the same temporal delay relative to the modes comprising the tidal stress \citep{EfroimskyMakarov13, makarov_efroimsky13}. The CPL model, on the other hand, is not supported by physical principles because it assumes a constant tidal response independent of the rotation frequency, which is not supported by laboratory and geophysical observations \citep{JacksonFaul10,jackson2005laboratory,Khan_etal18,Bagheri_etal19,LauFaul19}. The model employed here assigns a separate phase lag and amplitude response with appropriate adjustment for frequency dependence of the tidal mode through the viscoelastic model (Sundberg-Cooper), as discussed above.

Since the Pluto-Charon system is presently locked in a 1:1 spin-orbit resonance, backward integration of the orbit is not possible. Instead, several forward-in-time model computations are conducted starting from different initial conditions and searching for those model runs that lead to the observed orbital properties (Table~\ref{tab:orbital_params}). The heat arising from tidal dissipation is computed using Eq.~\eqref{eq5}, while the time evolution of the orbital parameters are computed via Eqs.~\eqref{eq1}-\eqref{eq4}. Finally, the initial inclination of Charon's orbit is assumed to be zero from the time of formation, which is a typical outcome of impact formation simulations \citep[e.g.,][]{ida_etal97,Canup05,Canupsalmon18,Citron_etal15}, as a result of which we do not track evolution in inclination.

\section{Results and discussion}\label{Numerical_results}
A ``nominal case" is chosen among the simulations that matches the present-day observed orbital parameters (Table~\ref{tab:orbital_params}) with interior properties and thermal parameters as given in Tables~\ref{tab:interior_props} and \ref{tab:thermal_parameters}, respectively. The initial orbital eccentricity, semi-major axis, and spin rates employed are $e$$\sim$0.4, $a$$\sim$0.65$a_{p}$, and $\sim$5 times the initial mean motion, corresponding to rotation period around 10~hr, respectively, where $a_p$ is the present-day observed semi-major axis. These values are based on the Pluto-Charon impact simulations of \citet{Canup05} and reflect the fact that in the post-collisional state, Charon's orbital eccentricity is high and both bodies commenced with spin rates that are higher than their initial orbital mean motion, indicative of a closer-in satellite. The choice of initial parameters will be discussed further in section~\ref{sec:discussion}.

\subsection{Tidal evolution}
\begin{figure} 
\begin{center} 
 \includegraphics[width=1.0\textwidth]{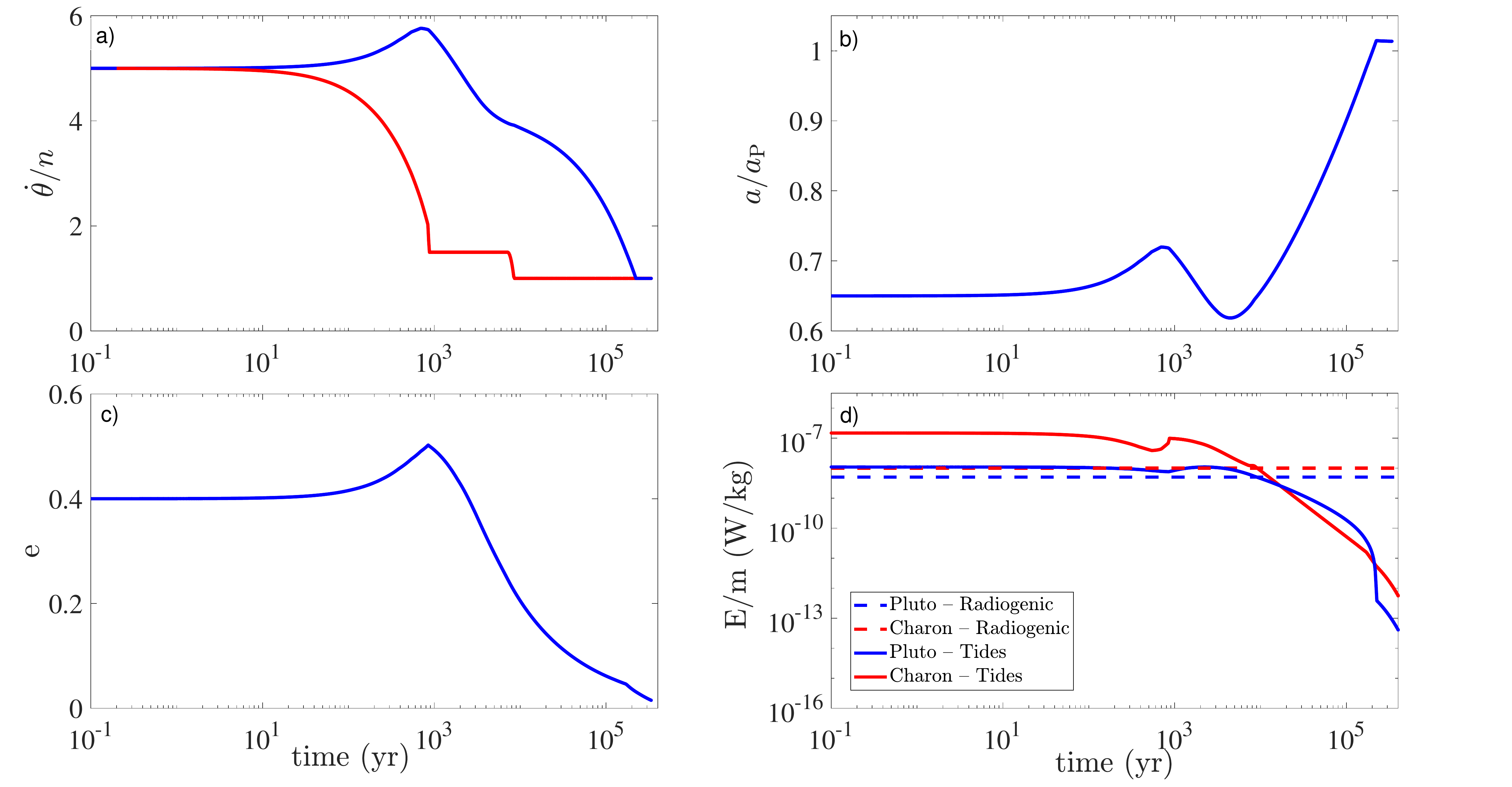}
\end{center}
\caption{Tidal evolution of the Pluto-Charon system. a) spin rates ($\dot\theta/n$), b) semi-major axis ($a/a_p$), c) eccentricity ($e)$, and d) generated heat (E) per mass (m) from tidal dissipation and decay of long-lived radioactive isotopes. $n$ and $a_p$ are the mean anomaly and the present-day distance between Pluto and Charon listed in Table~\ref{tab:orbital_params}.}
\label{fig:standard_evolution} 
\end{figure}

Figure~\ref{fig:standard_evolution} shows the evolution of the spin rates of the two bodies, semi-major axis of the orbit, orbital eccentricity, and the tidal heat generated in each of the objects. Figure~\ref{fig:standard_evolution}a shows that the tidal evolution of the binary system occurs very rapidly with the stable 1:1 spin-orbit resonance state ($\dot{\theta}/n = 1$) achieved in $\sim$$2\times 10{^5}$ years. Moreover, Charon's spin rate monotonically decreases until it falls into the 3:2 spin-orbit resonance ($\dot{\theta}/n = 1.5$) after $\sim$10$^4$ years. Once the eccentricity of the orbit has become sufficiently damped ($e<$0.2), the moon rapidly falls into the stable 1:1 spin-orbit resonance. A consequence of the capture into these resonances, is a slower orbital evolution of Charon and despinning rate of Pluto. Because of its smaller size, Charon reaches the 1:1 tidally-locked state much faster than Pluto. Pluto's spin rate, in contrast, increases in the beginning of the evolution but reverts once the satellite reaches the 3:2 resonance and only starts to properly decrease toward synchronization after Charon has entered the 1:1 resonance.

Figure~\ref{fig:standard_evolution}b shows the expansion of the semi-major axis. As illustrated in the plot, the orbit does not expand monotonically because of the aforementioned resonances. In fact, the orbit expands for a short period of time just before Charon's capture into the 3:2 spin-orbit resonance at which point the orbital separation decreases. Once the satellite falls out of this resonance and has reached the 1:1 spin-orbit resonance, Pluto is propelled toward synchronisation with the orbital motion $n$. At this point
the orbit begins to expand as angular momentum is transferred from Pluto’s super-synchronous spin rate into their mutual orbit, increasing the semi-major axis until the orbital evolution stabilizes at $a = a_p$ when the system has reached its final present-day dual-synchronous state.

Related to the spin rate and semi-major axis behaviour is the change in eccentricity.  Figure~\ref{fig:standard_evolution}c shows that the eccentricity of the orbit increases for a very short time in the beginning before Charon's fall into the 3:2 spin-orbit resonance, at which point, the eccentricity starts to dampen smoothly to 0, dictating the fall of Charon into the tidally-locked stated. A qualitatively similar behaviour in the orbital evolution was observed by \citet{Renaud_etal21}, but because of a different initial $\dot{\theta}/n$ and differences in planetary structure (physical properties and attenuation), the exact timing of the 3:2 and 1:1 resonance captures vary slightly in comparison to our results. 

Finally, Figure~\ref{fig:standard_evolution}d displays the heat generated in the two bodies due to both tidal and radiogenic heating. We can make two observations from this plot: 1) the amount of heat generated by tidal dissipation is higher in Charon than in Pluto, reaching $10^{-7}$~W/kg for Charon and $10^{-8}$~W/kg for Pluto, respectively, and is, relative to radiogenic heating, only of importance in the early stages of orbital evolution ($<10{^3}$~yr); 2) radiogenic heating approximately equals the maximum tidal heat generated in Pluto, and is approximately 10 times less in the case of Charon, but that the tidal heat source rapidly decreases as the eccentricity of the orbit diminishes; 3) radiogenic heating appears to govern the thermal evolution throughout most of the history of both bodies; and 4) as Pluto and Charon reach the dual-synchronous state, there is no longer any tidal force operating and tidal dissipation ceases to be an active heat source.
As the subsurface ocean evolves with time, the thermal evidence of despinning is no longer visible. Nonetheless, the tidal dissipation results in stresses within the body (not studied here) that can potentially  manifest in the form of surface features that survive until the present time, and can be used to constrain the past history of the objects \citep[e.g.,][]{Robuchon_Nimmo11,Rhoden20}.




\subsection{Evolution of thermal structure}
\begin{figure} 
\begin{center}
 \includegraphics[width=.7\textwidth]{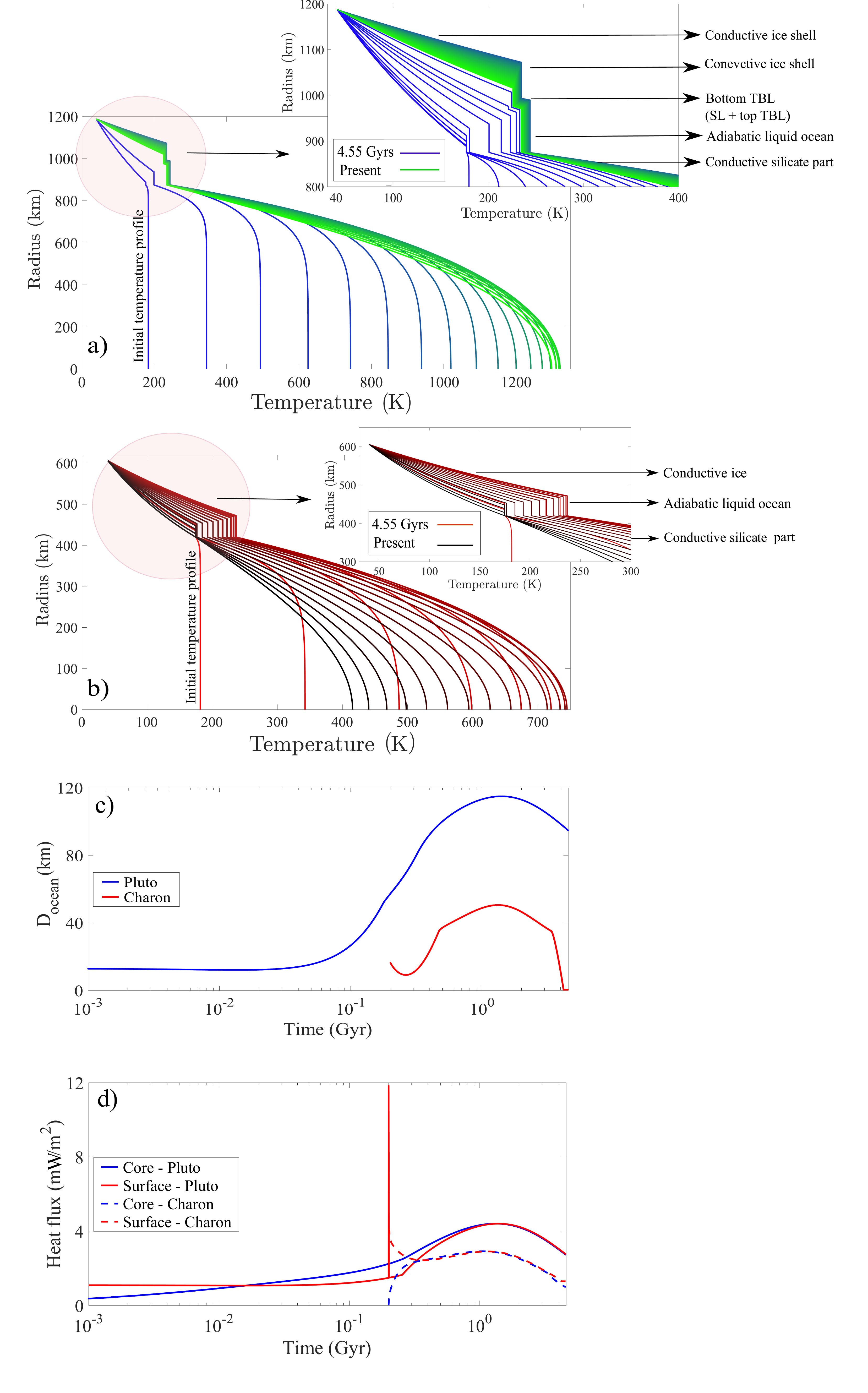}     
\end{center}
\caption{Temporal evolution of the thermal state of Pluto and Charon. (a) Radial temperature profiles of Pluto with each profile representing a time step of 200~Myrs (large panel) and zoom-in of water and ice layers (inset) in time steps of 15~Myrs. Note that in the inset, the convective ice shell starts by increasing (blue--dark green) in thickness but ends up by decreasing toward the end (light green). (b) as in (a) for Charon. Note that no convective ice shell develops on Charon. (c) Temporal evolution of the thickness of the liquid subsurface ocean on Pluto and Charon. (d) Temporal evolution of the surface heat flow on Pluto and Charon. The core heat flow has been scaled to that at the surface.}
\label{fig:temps}
\end{figure}

The thermal evolution of Pluto and Charon for the nominal case is shown in Figure~\ref{fig:temps}. Both nominal cases start with a conservative 15~km ocean thickness (other relevant parameters are compiled in Table~\ref{tab:thermal_parameters}), and develop oceans (see insets in Figure~\ref{fig:temps}a,b). In our computations. we focused on the time evolution of the ocean thickness and we found that convection facilitates growth of the ice layer because of the efficient removal of heat, which promotes the crystallization of the ocean. Our model runs actually show that while in most cases oceans develop on both bodies, these are unlikely to persist on Charon until the present. The exact conditions for this to occur depend on several parameters (mainly reference viscosity, radioactive element content, and impurity content) that will be discussed further in section~\ref{sec:discussion}. 

Figure~\ref{fig:temps} illustrates the thermal evolution of Pluto (Figure~\ref{fig:temps}a) and Charon (Figure~\ref{fig:temps}b) with time in terms of snapshots of the temperature profile (every 200~Myr). The central part made up of the silicate core heats up as a result of radioactive decay reaching temperatures of $\sim$1300~K (Pluto) and $\sim$800~K (Charon) after 4.5~Gyr of evolution. As core temperatures continue to rise, heat escapes and warms up the surrounding ice shell that softens, allowing for increased dissipation. Note that this is only really of importance in the very earliest stages of evolution, since tidal dissipation only dominates in the first 10$^4$~yr (Figure~\ref{fig:temps}d). Oceans grow progressively with time on both bodies as a consequence of the flow of heat from below and the accompanied melting of the ice layer. While the ocean on Pluto reaches a thickness of 100~km, it re-freezes entirely on Charon (Figure~\ref{fig:temps}c) on account of its size and therefore limited radiogenic  budget. As illustrated in Figure~\ref{fig:temps}a (inset), convection on Pluto starts operating in the ice shell after $\sim$300~Myr, resulting in a thermal structure composed of a hot thermal boundary layer (TBL) at the bottom, an adiabatic region in the middle of the shell, and a cold TBL and stagnant-lid at the top, 
as also observed elsewhere \citep[e.g.,][]{Mckinnon06,Deschamps2010,Robuchon_Nimmo11}. In comparison, because of Charon's smaller size and thinner ice shell, heat is effectively removed through conduction, and convection in the bottom part of the ice layer never sets in \citep[][]{Bierson_etal18, Nimmo_etal17}. In this particular simulation, Charon's subsurface ocean solidified completely $\sim$300~Myr ago, consistent with previous studies \citep[e.g.,][]{Bierson_etal18, Rhoden20} and the observation of extensional surface tectonic features (see section~\ref{interior_properties}). Up until about 3.5~Gyr on Charon, only H$_2$O crystallizes out of the ocean. As a consequence, the concentration of NH$_3$ in the ocean grows, reaching the eutectic at $\sim$3.5~Gyr, at which point we observe a change in slope in the crystallization behaviour of the ocean, which is due to the fact that both NH$_3$ and H$_2$O start to crystallize and the concentration of NH$_3$ in the remaining ocean no longer changes (on Pluto the eutectic is never reached). An interesting consequence of this result is that the bottom of the ice shell on Charon may be composed of a mix of H$_2$O and NH$_3$, assuming that the latter compound was present in the initial ocean.

Figure~\ref{fig:temps}d displays the evolution of surface and core heat flux (evaluated at the surface) on both bodies. As the central parts start to heat up, core heat flux increases, which continuously melts the initially conductive ice layer. Yet, as the temperature of the core increases, the viscosity of the ice shell decreases, which results in an increase of the convective heat flux (Eq.\eqref{eq:interior_viscosity}). A maximum amount of melting is reached after about 1~Gyr with ocean thicknesses of $\sim$110~km and $\sim$50~km for Pluto and Charon, respectively. At this point core and surface heat flux almost balance and this near-steady-state condition dominates the remainder (over $\sim$3.5~Gyr) of the evolution of Pluto and Charon as core heat flux continuously diminishes, because of insufficient radioactive heating, and the ocean layer refreezes. The spike in heat flux at 200~Myr is due to the tidal heating that is deposited in the ice layer as Pluto despins. This excess energy represents a transient episode, resulting in short-term melting of the ice that re-freezes as the transient fades away. 

\clearpage
\subsection{Influence of model parameters}\label{sec:discussion}

\subsubsection{Orbital parameters}
In the nominal case (Figure~\ref{fig:standard_evolution}), we relied on the outcome of impact models \citep[e.g.,][]{Canup_etal21} to guide our initial choice of orbital parameters. In the following, we briefly discuss the impact of these parameters on the tidal energy that is dissipated and the time for the Pluto-Charon system to evolve to the 1:1 synchronous state. Note that, unlike the thermal parameters, the effect of the orbital parameters, i.e., initial $a$, $e$, and $\dot{\theta}$ (hereinafter $a_0$, $e_0$, and ${\dot{\theta}}_0$), are not simply studied independently of each other. This is due to the fact that by varying one of the parameters, the present-day observed orbital configuration (Table~\ref{tab:orbital_params}) may not be achieved. For example, for an initial eccentricity of 0.2, with $a_0$ and ${\dot{\theta}}_0$ equal to the nominal case, the final 1:1 spin-orbit resonance state is reached at $a=1.2a_p$ (not shown).

In contrast to the nominal case, here we focus on an example, where the initial orbital parameters are closer to the present-day orbital values: $e_{0} = 0.06$, $\dot{\theta_{0}}/n_{0}=3$, and $a_0 = 0.7 a_p$ ($n_{0}$ is the initial mean motion). For this particular case, the maximum tidal heating rate in each of the two bodies is approximately one order of magnitude less than the nominal case, while the total time to reach the stable 1:1 spin-orbit state is not considerably different from that of the nominal case. We also considered the case where the initial spin rates of the two bodies were not equal. While such a case is plausible, we did not observe any significant difference in the main characteristics of the orbital evolution from the point of view of the time it takes for the Pluto-Charon system to arrive at its final state and the amount of tidally-dissipated heat.

As part of our simulations, we also considered cases where $a_0>a_p$, i.e., initial planet-satellite separation greater than the current distance. However, we found that none of them converged to the present-day orbit. This implies that the separation between the host and the satellite initially had to be smaller than the present-day semi-major axis and that the subsequent orbital evolution has acted to expand the orbit.  

We also investigated the effect of the frequency dependence ($\alpha$) on the orbital evolution, which mainly affects the heating rate caused by tidal dissipation. We considered two possibilities: $\alpha=0.1$ and $\alpha=0.4$, and found that in both cases the main features of the orbital evolution relative to the nominal case are not significantly different. For $\alpha=0.4$, the maximum heating rate can increase by up to one order of magnitude. Generally, we find that in all of our simulations the orbital evolution does not exceed $\sim$10$^6$~years and, relative to the contribution from radioactive decay, tidal heating has little affect on the final state of the Pluto-Charon system.





\subsubsection{Parameters governing the thermal evolution}
Our results show that for the nominal case, Pluto likely harbors a present-day ocean overlain by a conductive ice shell, whereas the ocean on Charon, although present up until $\sim$3.5~Gyr, has refrozen completely (Figure~\ref{fig:standard_evolution}c). Both of these results are in agreement with the observations by New Horizons of predominantly extensional and compressional tectonic features on Pluto and Charon, respectively. In the nominal case (Figure~\ref{fig:temps}), we fixed a number of key parameters that govern the thermal evolution of the two bodies. These include: reference viscosity, core size, initial thermal state, and ocean contaminants. In the following, we briefly discuss the influence of each of these parameters on the evolution of the ocean thickness.

\begin{figure} 
\hspace{-0mm}\includegraphics[width=1.05\textwidth]{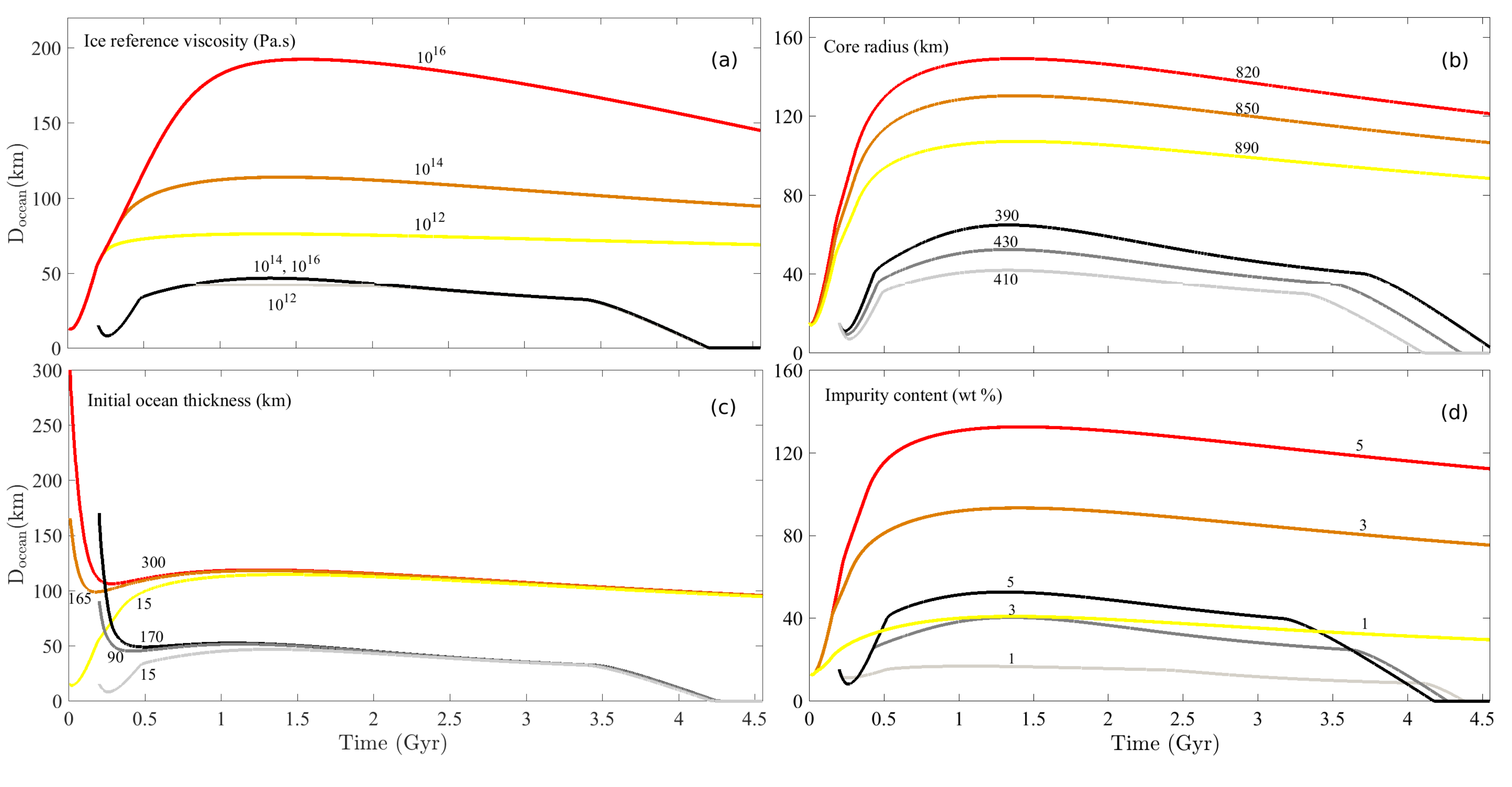}
\caption{Summary of thermal parameter study illustrating their influence on the evolution (ocean thickness) of the Pluto-Charon system.  (a) Ice reference viscosity; (b) Core radius; (c) Initial ocean thickness; and (d) Impurity content. In plots (a--d) Pluto is described by yellow, orange, and red lines, whereas Charon is delineated by light gray, gray, and black lines. The change in slope that is observed on all Charon-related curves toward the end of the evolution ($\sim$3.5~Gyr) is indicative of a
change in the crystallization behaviour of the ocean. See main text for more details.}
\label{fig:parameter_effect}
\end{figure}

\paragraph{Reference viscosity.}
Convection is very sensitive to viscosity and its variations with temperature. A high viscosity opposes the flow, reducing the strength of convection, while temperature-dependent viscosity triggers stagnant-lid convection, which alters the heat transfer. In our computations, the reference viscosity $\eta_{ref}$ controls the bulk viscosity of the ice shell (Eq.~\eqref{eq:interior_viscosity}).
Figure~\ref{fig:parameter_effect}a displays the evolution of ocean thickness on Pluto and Charon for several different values of reference viscosity ($\eta_{ref}$): 10$^{12}$, 10$^{14}$, and 10$^{16}$~Pa~s (all other relevant parameters are as in Table~\ref{tab:thermal_parameters}). Clearly, $\eta_{ref}$ has a strong impact on ocean thickness, in that larger viscosities generally result in a slower cooling as convection weakens, which increases the longevity and the thickness of the ocean layer. For all considered values of $\eta_{ref}$, Subsurface oceans on Pluto may survive until the present day, while the initially-formed oceans on Charon have all re-solidified. 

For Pluto in the low viscosity case, corresponding to $\eta_{ref} = 10^{12}$~Pa~s (yellow line), the ice shell melts until about 0.4~Gyr, leading to a $\sim$70-km thick ocean, which remains roughly constant for the remainder of the evolution. $\eta_{ref} = 10^{14}$~Pa~s (orange line) corresponds to the nominal case and has been discussed above. Convection was also observed for $\eta_{ref} = 10^{15}$~Pa~s (not shown). However,
for $\eta_{ref} = 10^{16}$~Pa~s (red line), convection never commences, and melting of the ice sheet is dominant in the first $\sim$1~Gyr. This results in the conductive shell becoming progressively thinner as the core continues to heat up, with the ocean reaching a maximum thickness of $\sim$200~km. With core heat flux subsequently declining as radioactive heating diminishes, the ocean starts to re-freeze and the ice shell thickens. Because the ocean re-freezes at a relatively slow rate, a subsurface ocean on Pluto is able to persist to the present day, which occurs in all of the considered cases. 
On Charon, convection is unable to operate, because 1) the ice shell is not thick enough and 2) the effect of temperature-dependent viscosity is too strong, and an ocean, while present for $\sim$3~Gyr, is unable to remain liquid to the present. Due to the absence of convection on Charon the influence of viscosity is considerably smaller than in the case of Pluto.

\paragraph{Core size and initial ocean/ice layer thickness.}
Varying core size implies changing the bulk radioactive element content and, through that, core heat flow. We varied core size on Pluto (820--870~km) and Charon (390--430~km) and adjusted core density (3--3.6~g/cm$^3$) to ensure that total mass of the body remains unchanged. All other relevant parameters are as in Table~\ref{tab:thermal_parameters}. The results are shown in Figure~\ref{fig:parameter_effect}b and illustrate, as above, that oceans on Pluto may still exist today, while oceans on Charon have all re-solidified, with the exception of the case of a small (390~km) core that predicts a thin ($\leq$5~km) subsurface ocean on Charon at the present day.
Moreover, the changes in core size result in subsurface oceans that vary in thickness by $\sim$30~km and $\sim$20~km on Pluto and Charon, respectively. Again, convection on Charon is generally inhibited by the presence of a smaller hydrosphere.

We further investigated the effect of initial ocean/ice layer thickness (Figure~\ref{fig:parameter_effect}c), which, as expected, shows the largest differences in the early stages of evolution, where initially thick ice layers (thin ocean layers) follow the path described earlier (Figure~\ref{fig:parameter_effect}a) for the low reference viscosity cases ($\eta_{ref}$=10$^{12}$~Pa~s), whereas thin ice layers (thick ocean layers) immediately start off by freezing until a quasi-steady-state condition is reached, where the heat entering from below is conductively removed through the ice layer. Independently of initial ocean/ice layer thickness, Figure~\ref{fig:parameter_effect}c shows that subsurface oceans on Pluto and Charon reach thicknesses $\sim$100~km and 0~km, respectively. These results show that thermal evolution completely removes the signature of the initial ice shell. In summary, the initial ocean/ice layer thickness is less important in the context of the long-term evolution of Pluto and Charon.

\paragraph{Impurity content.}
The presence of impurity species reduces the temperature at the bottom of the ice shell, which increases the bulk viscosity and thus reduces the vigor of convection. To evaluate the importance of this effect on the evolution of Pluto and Charon, we vary the initial fraction of ammonia in the range 1--5~wt\% (all other relevant parameters are as in Table~\ref{tab:thermal_parameters}). The results are shown in Figure~\ref{fig:parameter_effect}d for both bodies. Two conclusions may be drawn: 1) regardless of impurity content, a subsurface ocean is always present on Pluto at the present day, and 2) oceans also form on Charon, but completely solidify after $\sim$3.5--4~Gyr. Between the various cases, ocean thicknesses on Pluto and Charon range from $\sim$10--50~km and $\sim$40--120~km, respectively. Convection only happens for NH$_3$=1 and 3~wt\%, respectively, whereas for NH$_3$=5~wt\%, the ice layer remains conductive. Overall, convection within Charon appears to be difficult to set up and to maintain. For instance, even for NH$_3$=1~wt\%, convection is only operative between $\sim$0.25 and 2.4~Gyr. With reference viscosity, impurity content has the largest influence on final ocean thickness, in particular for Pluto, because of its impact on the vigor of convection.

\section{Summary and outlook}\label{sec:conclusion}
In this work we have implemented a comprehensive semi-analytical tidal model based on the latest advances in tidal theory that considers dual dissipation in a binary system and invokes an appropriate viscoelastic rheology (Sundberg–Cooper) for proper computation of the tidal response functions. Because the spin rate of a planet can impart orbital changes that lead, via dissipation, to changes in the thermal state of the other body, a thermal-orbital feedback exists that calls for a joint approach. Consequently, here we studied the combined tidal and thermal evolution of the Pluto-Charon system with particular emphasis on the possibility for present-day subsurface oceans on Pluto and Charon. 

We found that a subsurface ocean always develops and remains liquid to the present day on Pluto. Subsurface oceans on Charon also developed in all of the studied cases in its past (up until $\sim$3.5~Gyr), but because the ice layer is relatively thin, convection is unable to operate effectively, as a consequence of which the ocean re-freezes between 0--500~Myr ago. These observations were found to be robust across an entire range of orbital and thermal model parameter values that were considered in this study. Based on the parameter analysis, the present-day ocean on Pluto ranges in thickness from 40~km to 150~km. Observations of extensional and compressional surface tectonics on Charon and Pluto, respectively, made by New Horizons supports this. While we applied our thermal-tidal model to the Pluto–Charon system, it stands to reason that it can be applied to any satellite orbiting a central body (e.g., Neptune-Triton, Earth-Moon, or an exoplanet system). 

When viewed over a time span of $\sim$4.5~Gyr, the tidal energy that is dissipated in both bodies as the orbits evolve toward synchronization, including higher spin-orbit resonances, is less significant relative to the heat emanating from the decay of long-lived radioactive isotopes. Generally, the tidally-deposited energy from despinning is only of relevance in the early stages of the evolution of the Pluto-Charon system. In spite of the fact that tidal heating plays a less important role, our calculations nevertheless show that during the short period that tides were active, the amount of energy released by tides was much higher than that emanating from radioactive decay. This is of particular importance in the case of the tidally-active Jovian and Saturnian satellites \citep{asphaug_Reufer13,shoji_etal14,Neveu_Rhoden19,bierson_etal21,lainey_etal20,tyler_etal15} and exoplanetary systems \citep{miller_etal09,Shoji_Kurita14,driscoll_Barnes15,Dobos_etal19}.

In the context of tidally active and potentially habitable worlds, heating due to tidal dissipation in the oceans, i.e., from friction generated at the ice-water and/or water-solid mantle interfaces, may be important \citep{tyler08,Beuthe19}. On Earth, for example, 
the dissipation due to tidally-induced motion in the oceans is greater than that arising from the tidal deformation of the solid-body interior \citep{egbert_Ray03}. In the case of subsurface oceans, resonantly forced tidal states in the ocean with highly elevated power levels may even dominate the viscoelastic tidal dissipation in the solid parts of the body \citep{tyler20}. Tidal dissipation in the liquid ocean is a complex process and depends on parameters such as depth, composition, and the topography of the ice and ocean bottom \citep{tyler08} that are currently unknown. Here we have excluded the possibility for this additional heat source, but we may observe that had we incorporated the effect of ocean tidal heating, this would act to shorten the duration for complete synchronization. However, as tidal heating is only really significant very early on, the potential effect of considering ocean tides would not lead to any change in the long-term evolution. From the point of view of tidal heating and evolution timescales, our results represent a lower and an upper bound, respectively. 
 
Because direct information on the interiors of icy ocean worlds is limited, their astrobiological potential is difficult to assess in detail. For this, the key parameters relating to habitability, which include ocean depth, temperature, and chemistry, among others, need to be well-established. Given the high resolving power of geophysical methods, particularly of seismology, in situ geophysical exploration of icy planets and satellites is clearly the means to move beyond the current impasse as pointed out in numerous studies \citep[e.g.,][]{KovachChyba01,lee_etal03,Panning_etal06,Panning_etal18,vance_etal18,Stahler_etal18,Maguire_etal21}.
The successful Mars Insight mission has paved the way for single-station emplacement and sounding of icy worlds, in that InSight has shown that with a single station, marsquakes can be recorded, located, and used to sound its interior \citep{Knapmeyer_etal21,Khan_etal21,Staehler_etal21}.
In this context, the proposed Europa \citep{pappalardo_etal13} and the selected Titan relocatable lander ``Dragonfly" \citep{Barnes_etal21} missions both include a single-station seismometer in their proposed scientific payload that should enable better characterization of ocean thickness, for example.

\section{Acknowledgements}
This work was supported by a grant from the Swiss National Science Foundation (project 172508). We would like to acknowledge Joe Renaud and Michael Efroimsky for informed discussions on tidal evolution and dissipation.

\bibliographystyle{apalike}
\bibliography{Chapter_3}

\begin{appendix}

\section{Sundberg-Cooper viscoelastic model}\label{App:A}
The compliance or creep function associated with the Sundberg-Cooper model is \citep{SundbergCooper10}:
\begin{equation}
  J(\omega) =  J_U - \frac{i}{\eta \omega} + \frac{j J_R}{i-J_R \eta_1 \omega} + J_U(i J_U\eta \omega)^{-\alpha} \alpha!,
\end{equation}
where $J_U$ and $J_R$ are unrelaxed and relaxed shear moduli, respectively, $\eta_1$ is the Kelvin-Voigt viscosity, $\eta$ is viscosity ($\eta$ for the various layers is listed in Table~\ref{tab:thermal_parameters}), $\omega$ is frequency, and $\alpha$ is the frequency exponent as defined in the main text (see section~\ref{sec:viscoelast}).
The frequency-dependent complex shear modulus is given by
\begin{equation}
\Re[G(\omega)] = \frac{1}{J_U  H(\omega)} \bigg[ \big( [J_U \eta \omega ]^{\alpha}C +1\big) \lambda + \frac{J_U}{J_R}\bigg],
\end{equation}
\begin{equation}
\Im[G(\omega)] =\frac{\eta  \omega}{H(\omega)} \bigg[(J_U \eta \omega)^{-1-\alpha}  S\lambda+\frac{\eta_1}{\eta} + \big(\frac{\eta_1}{\eta}\big)^2 + (\eta \omega J_R)^{-2}\bigg],
\end{equation}
where $\lambda$ is the Burgers coefficient, $S$ and $C$ are constants, and the function $H(\omega)$ is defined as
\begin{equation}
H(\omega) = \frac{1}{(J_R\eta \omega)^2} + \lambda + 2 \frac{J_U}{J_R} + \bigg( \frac{\eta_1}{\eta} + 1 \bigg)^2 + \lambda (J_U \eta \omega  )^{-2\alpha} (a!)^2 + 2(J_U\eta \omega  )\bigg[S \bigg(J_U \eta_1 \omega \big(1+\frac{\eta_1}{\eta} \big) +\frac{J_U}{J_R (J_R \eta \omega )} \bigg) + C \big(\lambda + \frac{J_U}{J_R} \big)\bigg].
\end{equation}
The parameters and constants used in the above expressions are given in Table~\ref{tab:viscoparameters}.
The intrinsic shear quality factor $Q^{-1}$, which is not to be confused with the $Q^{-1}$ associated with global tidal dissipation (Eq.~\ref{dissip9}), is a measure of dissipation and can be computed from
\begin{equation}
    Q^{-1} = \frac{\Im[G(\omega)]}{\Re[G(\omega)]}.
\end{equation}

\begin{table}
\begin{center}
\begin{tabular}{lc|c}
\hline
\textbf{Parameter (Unit)}           &    \textbf{Symbol}          & \textbf{Value/expression}      \\ \hline
\multicolumn{1}{l}{ \quad Constant }   & S & $\alpha! \sin(\alpha \pi/2)$ \\
\multicolumn{1}{l}{ \quad Constant } & C   & $\alpha ! \cos(\alpha \pi/2)$   \\ 
\multicolumn{1}{l}{ \quad Burgers Coefficient} & $\lambda$  & $(J_U \eta_1 \omega)^2 + (J_U/J_R)^2$       \\ 
\multicolumn{1}{l}{ \quad Relaxed compliance $\rm (1/Pa.s)$} & $J_R$ & $0.2J_U$       \\ 
\multicolumn{1}{l}{ \quad Voigt-Kelvin Viscosity $\rm (Pa.s)$}  & $\eta_1$ &    $0.02\eta$   \\ 
\hline \hline
\end{tabular}
\caption{Parameters associated with the Sundberg-Cooper viscoelastic model.}
\label{tab:viscoparameters}
\end{center}
\end{table}
\section{Orbital evolution theory}\label{App:B}
In what follows, we rely on \cite{Bagheri_etal21} for $da/dt$ and $de/dt$, and augment these here with full expressions for $d\dot\theta/dt$, $d E/dt$, and the case of non-synchronous rotation. The time evolution of each orbital parameter of the two-body system can be cast as
\begin{equation}
  \bigg( \frac{dx}{dt} \bigg) = \bigg( \frac{dx}{dt} \bigg)_{\text{primary}} + \bigg( \frac{dx}{dt} \bigg)_{\text{secondary}}, 
\label{dxdtgeneral}
\end{equation}
where $x$ is either $a$ or $e$. The two terms refer to the tides in the planet and the tides in the satellite, respectively. Because of the spherical shape of both Pluto and Charon, we can omit the contribution from libration. 
In what follows, we provide only the main formulae for the orbital evolution. 

The expression for the semi-major axis rate is 
\begin{equation}\label{closeddadt}
\begin{aligned}
 \frac{da}{dt} = -2 a n \sum_{l=2}^{\infty} \sum_{m=0}^l \frac{(l-m)!}{(l+m)!} (2-\delta_{0m})\sum_{p=0}^{l}\sum_{q=-\infty}^{\infty} G_{lpq}^2(e) (l-2p+q) \\
 \bigg[ \bigg( \frac{R}{a} \bigg)^{2l+1} \frac{M'}{M} F_{lmp}^2(i)K_l(\beta) + \bigg( \frac{R'}{a} \bigg)^{2l+1} \frac{M}{M'}  F_{lmp}^2(i')K_l(\beta')\bigg],
\end{aligned}
\end{equation}
where $M$ and $M^{\prime}$ are the planet and satellite masses, $R$ and $R^{'}$ are their radii, $K_l = k_l/Q_l$ and $K_l^{\,\prime}=k_l^{\,\prime}/Q_l^{\,\prime}$ are the planet and satellite quality functions, $G(e)$ and $F(i)$ are the eccentricity and inclination functions, respectively, $\beta$ is the tidal mode, $n$ is the mean motion, and all the other variables are as defined in Eqs.~\eqref{eq1}--\eqref{eq3}. Since the inclinations of the orbits remain small, only $F_{201}$ and $F_{220}$ are relevant and are equal to $\frac{1}{2}$ and $3$, respectively. 
Similarly to $da/dt$, the general expression for the eccentricity rate is
\begin{equation}\label{closeddedt}
\begin{aligned}
& \frac{de}{dt} = -\frac{1-e^2}{e}  \frac{n}{MM'} \sum_{l=2}^{\infty} \sum_{m=0}^l \frac{(l-m)!}{(l+m)!} (2-\delta_{0m})  \sum_{p=0}^{l}\sum_{q=-\infty}^{\infty} G_{lpq}^2(e) (l-2p+q) \\
& \bigg[ \bigg( \frac{R}{a} \bigg)^{2l+1} \frac{M'}{M} F_{lmp}^2(i)K_l(\beta) + \bigg(  \frac{R'}{a} \bigg)^{2l+1} \frac{M}{M'} F_{lmp}^2(i')K'_l(\beta') \bigg] + \\
& -\frac{\sqrt{1-e^2}}{e} \frac{n}{MM'} \sum_{l=2}^{\infty} \sum_{m=0}^l \frac{(l-m)!}{(l+m)!} (2-\delta_{0m})\sum_{p=0}^{l}\sum_{q=-\infty}^{\infty} G_{lpq}^2(e) (l-2p)  \\
&  \bigg[ \bigg( \frac{R}{a} \bigg)^{2l+1} \frac{M'}{M} F_{lmp}^2(i)K_l(\beta) + \bigg( \frac{R'}{a} \bigg)^{2l+1} \frac{M}{M'} F_{lmp}^2(i')K'_l(\beta')  \bigg].
\end{aligned}
\end{equation}
Similarly, equations for the tidal heating and the tidal torque are given as:
\begin{equation}\label{closeddenergydt}
\begin{aligned}
 \bigg(\frac{dE}{dt}\bigg)_{\text{primary}} = \frac{GM'^2}{a} \sum_{l=2}^{\infty} \bigg(\frac{R}{a}\bigg)^{2l+1} \sum_{m=0}^l \frac{(l-m)!}{(l+m)!} (2-\delta_{0m})\sum_{p=0}^{l}\sum_{q=-\infty}^{\infty}G_{lpq}^2(e) F_{lmp}^2(i)K_l(\beta') \beta',
\end{aligned}
\end{equation}

\begin{equation}\label{closeddthetadt}
\begin{aligned}
\bigg(\frac{d\dot\theta}{dt}\bigg)_{\text{primary}}  = \frac{GM'^2}{C a} \sum_{l=2}^{\infty} \bigg(\frac{R}{a}\bigg)^{2l+1} \sum_{m=0}^l \frac{(l-m)!}{(l+m)!} (2-\delta_{0m})\sum_{p=0}^{l}\sum_{q=-\infty}^{\infty}G_{lpq}^2(e) F_{lmp}^2(i)m K_l(\beta'),
\end{aligned}
\end{equation}
where $C$ is the polar moment of inertia of the planet. Similar expressions are obtained for the satellite are obtained by replacing $M$ with $M^{\prime}$ and $C$ by $C^{\prime}$, where $C^{\prime}$ is the polar moment of inertia of the secondary.

Due to low convergence of the series and the relatively high eccentricities found in this study, we have to include higher-order terms to ensure precision of our results and stability of integration for high eccentricities. 
In Eq.~\eqref{dxdtgeneral}, the contribution of tides raised by the satellite in the planet is
 \begin{equation}\label{dadtmainplanet}
 \begin{aligned}
 & \bigg( \frac{da}{dt} \bigg)_{\text{primary}} = n\bigg( \frac{R^5}{a^4} \bigg) \bigg( \frac{M^{\,\prime}}{M} \bigg) \times {\cal{F}}(K_l,\,\dot\theta,\,n,\,e)~~. \\
 \end{aligned}
 \end{equation}
The contribution due to the tides raised by the planet in the satellite looks similar
\begin{equation}
\bigg( \frac{da}{dt} \bigg)_{\text{secondary}} = n\bigg( \frac{{R^{\,\prime}}^5}{a^4} \bigg) \bigg( \frac{M}{M^{\,\prime}} \bigg) \times {\cal{F}} ( K^{\,\prime}_l,\,\dot \theta^{\,\prime},\,n,\,e)~~.
\label{dadtmainsatellite}
\end{equation}
Here $\,{\cal{F}}\,$ is a function of the eccentricity, the mean motion, the satellite spin rate ($\dot\theta'$) and its tidal response
\begin{equation}\label{fcoeffs}
	{\cal{F}} \left( K_l', \dot\theta', n, e\right) =\sum\limits_{i=0}^9e^{2i}\left(\sum\limits_{j=-7}^{-1}K_l'\left(jn-2\dot\theta'\right)\varphi_{j}^{2i}+\sum\limits_{j=1}^{11}K_l'\left(jn-2\dot\theta'\right)\varphi_j^{2i}+\sum\limits_{j=1}^9K_l'(jn) \hat \varphi_j^{2i}\right).
\end{equation}
The coefficients $\varphi_j^{2i}$ and $\hat \varphi_j^{2i}$ of the series are tabulated in Supplementary Tables~3 and 4. Note that in these tables, the terms of the series that are not specifically mentioned are equal to zero.
The above equations have been derived for the general case, i.e., with neither of the bodies assumed synchronous. In the specific situation of a synchronised moon, we have $\,\dot \theta^{\,\prime} = n\,$, wherefore the semi-diurnal term in equation \ref{dadtmainsatellite} vanishes: $\,K_l(2n-2 \dot \theta^{\,\prime})\,=\,0\,$. In the contribution from the planet, the semi-diurnal term vanishes when the satellite is at the synchronous orbit, i.e., when $\,n=\dot \theta'\,$.

Similarly, to compute the eccentricity evolution, we write down all the inputs entering Eq.~\eqref{closeddedt}. The input generated by the tides in the planet is
\begin{equation}\label{dedtmainplanet}
\begin{aligned}
\bigg( \frac{de}{dt} \bigg)_{\text{primary}} = -n \frac{M'}{M} \bigg( \frac{R}{a} \bigg)^5 \times {\cal{L}}(K_l,\dot\theta,n,e)~~,
\end{aligned}
\end{equation}
while the input from the tides in the satellite reads as
 \begin{equation}
 \begin{aligned}
 \bigg(\frac{de}{dt} \bigg)_{\text{secondary}} = -n \frac{M}{M'} \bigg( \frac{R'}{a} \bigg)^5 \times {\cal{L}}(K^{\,\prime}_l,\dot \theta^{\,\prime},n,e)~~,
 \end{aligned}
 \label{mainsatdedt}
 \end{equation}
where the function $\cal{L}$ is defined as
 \begin{equation}
	{\cal L}(K_l,\dot\theta,n,e) = \sum\limits_{i=1}^9e^{2i-1}\left(\sum\limits_{j=-7}^{-1}K_l\left(jn-2\dot\theta\right)\lambda_{j}^{2i-1}+\sum\limits_{j=1}^{11}K_l\left(jn-2\dot\theta\right)\lambda_j^{2i-1}+\sum\limits_{j=1}^9K_l(jn)\hat \lambda_j^{2i-1}\right)~~.
	\label{eq:lcoeffs}
\end{equation}
The coefficients $\lambda_j^{2i-1}$ and $\hat \lambda_j^{2i-1}$ are tabulated in Supplementary Table~6 and 7.
Note that, similarly to $da/dt$, the expression is general, in that neither of the bodies is {\it{a priori}} assumed synchronous. For the synchronised case, the term associated with the semi-diurnal tide in equation \eqref{dedtmainplanet} vanishes.

To compute the time evolution of the tidally dissipated energy, we expand Eq.~\eqref{closeddenergydt} as follows 
\begin{equation}\label{denergydtmainplanet}
\begin{aligned}
\bigg( \frac{dE}{dt} \bigg)_{\text{primary}} =  \frac{GM'^2}{a} \bigg( \frac{R}{a} \bigg)^5 \times {\cal{H}}(K_l,\dot\theta,n,e)~~,
\end{aligned}
\end{equation}
The expression for ${\cal{H}}$ is given by 
\begin{equation}
	{\cal{H}}(K_l,\dot\theta,n,e) = \sum\limits_{i=0}^9e^{2i}\left[\sum\limits_{j=-7}^{-1}\CC_j^{2i}|jn-2\dot \theta|K_l\left(|jn-2\dot \theta|\right) +\sum\limits_{j=1}^{11}\CC_j^{2i}|jn-2\dot \theta|K_l\left(|jn-2\dot \theta|\right)+\sum\limits_{j=1}^9jn\CCp_j^{2i}K_l(jn)\right]
\end{equation}

Finally, the expression for the time evolution of the spin rate, Eq.~\eqref{closeddthetadt} can be expanded as 
\begin{equation}\label{dthetadtmainplanet}
\begin{aligned}
\bigg( \frac{d\dot{\theta}}{dt} \bigg)_{\text{primary}} = \frac{GM'}{Ca} \bigg( \frac{R}{a} \bigg)^5 \times {\cal{G}}(K_l,\dot\theta,n,e)~~,
\end{aligned}
\end{equation}
The expression ${\cal{G}}$ is
\begin{equation}
	{\cal{G}}(K_l,\dot\theta,n,e) = \sum\limits_{i=0}^9e^{2i}\left[\sum\limits_{j=-7}^{-1}\chi_j^{2i}K_l\left(jn-2\dot \theta\right) +\sum\limits_{j=1}^{11}\chi_j^{2i}K_l\left(jn-2\dot \theta\right)\right]
\end{equation}
\end{appendix}
\clearpage
\section*{Supplementary material: Tidal coefficients}\label{Supp:1}

This section summarizes all the tidal coefficients that are required in the development of the orbital theory as formulated in the Methods section of the main text, i.e., Eq.~\eqref{eq2}--\eqref{eq5}.

\clearpage
\begin{table}\caption{$\varphi_j^{2i}$ }
			\hspace{50mm}
	\begin{tabular}{c|ccccccc}
		\multicolumn{5}{c}{\framebox{$ \varphi_{j}^{2i}$ }}    
		\\
		$_j \backslash ^{2i}$ & $ 0  $ & $ 2  $             & $ 4  $              & $ 6  $                    \\ \hline
		$ -1 $             & $0 $   & $0 $               & $0 $                & $ \frac{1}{1536} $        \\
		$ 1 $              & $0 $   & $ -\frac{3}{8} $   & $ \frac{3}{32} $    & $ -\frac{13}{512} $       \\
		$ 2 $              & $ -3 $ & $ 15 $             & $ -\frac{189}{8} $  & $ \frac{155}{12} $        \\
		$ 3 $              & $0 $   & $ -\frac{441}{8} $ & $ \frac{7749}{32} $ & $ -\frac{197775}{512} $   \\
		$ 4 $              & $0 $   & $0 $               & $ -\frac{867}{2} $  & $ 1955 $                  \\
		$ 5 $              & $0 $   & $0 $               & $0 $                & $ -\frac{3570125}{1536} $ \\
		\hline
		\end{tabular}\\
		\end{table}
\begin{table}
\begin{tabular}{c|ccccccc}
	\multicolumn{7}{c}{\framebox{$ \varphi_{j}^{2i}$ }}    
	\\
	$_j \backslash {2i}$ & $ 8  $                      & $ 10  $                           & $ 12  $                           & $ 14  $                                   & $ 16  $                                   & $ 18  $                                                \\ \hline
	$ -7 $             & $0 $                        & $0 $                              & $0 $                              & $0 $                                      & $0 $                                      & $ \frac{232630513987207}{469654673817600} $            \\
	$ -6 $             & $0 $                        & $0 $                              & $0 $                              & $0 $                                      & $ \frac{4782969}{20070400} $              & $ \frac{1594323}{20070400} $                           \\
	$ -5 $             & $0 $                        & $0 $                              & $0 $                              & $ \frac{1220703125}{11098128384} $        & $ \frac{13427734375}{177570054144} $      & $ \frac{1619873046875}{14611478740992} $               \\
	$ -4 $             & $0 $                        & $0 $                              & $ \frac{32}{675} $                & $ \frac{32}{675} $                        & $ \frac{92}{1575} $                       & $ \frac{7808}{127575} $                                \\
	$ -3 $             & $0 $                        & $ \frac{59049}{3276800} $         & $ \frac{59049}{2621440} $         & $ \frac{38558997}{1468006400} $           & $ \frac{18482337}{671088640} $            & $ \frac{72655602021}{2630667468800} $                  \\
	$ -2 $             & $ \frac{1}{192} $           & $ \frac{7}{960} $                 & $ \frac{949}{115200} $            & $ \frac{10193}{1209600} $                 & $ \frac{127453}{15482880} $               & $ \frac{781813}{99532800} $                            \\
\hline
	\end{tabular}\label{tab:coeffFij}
	\end{table}
	
	\clearpage
	\begin{table}\caption{$ \hat  \varphi_{j}^{2i}$ }
	\hspace{60mm}
	\begin{tabular} {c|ccccccc}
		\multicolumn{5}{c}{\framebox{$\hat \varphi_{j}^{2i} $}}                                                   
		\\
		$_j \backslash ^{2i}$ & $ 0 $ & $ 2 $            & $ 4 $              & $ 6 $                 \\ \hline
		$ 1 $              & $0$   & $ -\frac{9}{4} $ & $ -\frac{81}{16} $ & $ -\frac{2295}{256} $ \\
		$ 2 $              & $0$   & $0$              & $ -\frac{81}{8} $  & $ -\frac{63}{4} $     \\
		$ 3 $              & $0$   & $0$              & $0$                & $ -\frac{8427}{256} $ \\
	\hline
	\end{tabular}
	\end{table}
\vspace{15mm}
\vspace{-20mm}
 
 \begin{table}
  \begin{tabular}{c|ccccccc}

	\multicolumn{7}{c}{\framebox{$ \hat  \varphi_{j}^{2i}$ }}                                                                                                                                                                                                        \\
	$_j \backslash ^{2i}$ & $ 8 $                   & $ 10 $                       & $ 12 $                        & $ 14 $                               & $ 16 $                                 & $ 18 $                                       \\ \hline
	$ 1 $              & $ -\frac{28403}{2048} $ & $ -\frac{3240741}{163840} $  & $ -\frac{87480539}{3276800} $ & $ -\frac{76230921863}{2202009600} $  & $ -\frac{143189681823}{3288334336} $   & $ -\frac{3798168072196597}{71028021657600} $ \\
	$ 2 $              & $ -\frac{1661}{64} $    & $ -\frac{12027}{320} $       & $ -\frac{525521}{10240} $     & $ -\frac{13519469}{201600} $         & $ -\frac{729537581}{8601600} $         & $ -\frac{56670971611}{541900800} $           \\
	$ 3 $              & $ -\frac{62487}{2048} $ & $ -\frac{18059643}{327680} $ & $ -\frac{96458913}{1310720} $ & $ -\frac{10219795947}{104857600} $   & $ -\frac{1455222699297}{11744051200} $ & $ -\frac{28832512267869}{187904819200} $     \\
	$ 4 $              & $ -\frac{5929}{64} $    & $ -\frac{9933}{320} $        & $ -\frac{708871}{6400} $      & $ -\frac{1736021}{14400} $           & $ -\frac{27865091}{172032} $           & $ -\frac{2578148063}{12902400} $             \\
	$ 5 $              & $0$                     & $ -\frac{15717645}{65536} $  & $ \frac{14736585}{262144} $   & $ -\frac{67620980465}{264241152} $   & $ -\frac{212874343975}{1409286144} $   & $ -\frac{13593131130225}{52613349376} $      \\
	$ 6 $              & $0$                     & $0$                          & $ -\frac{30089667}{51200} $   & $ \frac{20778687}{44800} $           & $ -\frac{592612065}{802816} $          & $ -\frac{124700799}{20070400} $              \\
\hline
	\end{tabular}\label{tab:coeffFhatij}
 \end{table}
\vspace{-5mm}
 \begin{table}\caption{$\gamma_{0}^{j,s}$}
\begin{tabular} {c|ccccccccc}
	\multicolumn{10}{c}{\framebox{$\gamma_{0}^{j,s}$}}                                                                                                                                                                                           \\
	$_j \backslash ^s$ & $  2 $  & $  4 $            & $  6 $            & $  8 $             & $  10 $             & $  12 $               & $  14 $                  & $  16 $                   & $  18 $                       \\ \hline
	$ 1 $              & $  -3 $ & $  3 $            & $  -\frac{5}{4} $ & $  \frac{7}{24} $  & $  -\frac{7}{160} $ & $  \frac{11}{2400} $  & $  -\frac{143}{403200} $ & $  \frac{143}{6773760} $  & $  -\frac{2431}{2438553600} $ \\
	$ 2 $              & $0 $    & $  -\frac{3}{2} $ & $  1 $            & $  -\frac{7}{24} $ & $  \frac{1}{20} $   & $  -\frac{11}{1920} $ & $  \frac{143}{302400} $  & $  -\frac{143}{4838400} $ & $  \frac{221}{152409600} $    \\
	$ 3 $              & $0 $    & $0 $              & $  -\frac{1}{4} $ & $  \frac{1}{8} $   & $  -\frac{9}{320} $ & $  \frac{11}{2880} $  & $  -\frac{143}{403200} $ & $  \frac{13}{537600} $    & $  -\frac{221}{174182400} $   \\
	$ 4 $              & $0 $    & $0 $              & $0 $              & $  -\frac{1}{48} $ & $  \frac{1}{120} $  & $  -\frac{11}{7200} $ & $  \frac{13}{75600} $    & $  -\frac{13}{967680} $   & $  \frac{17}{21772800} $      \\
	$ 5 $              & $0 $    & $0 $              & $0 $              & $0 $               & $  -\frac{1}{960} $ & $  \frac{1}{2880} $   & $  -\frac{13}{241920} $  & $  \frac{1}{193536} $     & $  -\frac{17}{48771072} $     \\
	$ 6 $              & $0 $    & $0 $              & $0 $              & $0 $               & $0 $                & $  -\frac{1}{28800} $ & $  \frac{1}{100800} $    & $  -\frac{1}{752640} $    & $  \frac{17}{152409600} $     \\
\hline
 \end{tabular}
 \end{table}
 \begin{tabular} {c|ccccccccc}
	\multicolumn{10}{c}{\framebox{$ \gamma_{1}^{j,s} $}}                                                                                                                                                  \\
	$_j \backslash ^s$ & $  1 $  & $  3 $   & $  5 $            & $  7 $              & $  9 $               & $  11 $             & $  13 $                   & $  15 $                    & $  17 $                    \\ \hline
	$ 1 $              & $  30 $ & $  -42 $ & $  \frac{45}{2} $ & $  -\frac{77}{12} $ & $  \frac{91}{80} $   & $  -\frac{11}{80} $ & $  \frac{2431}{201600} $  & $  -\frac{2717}{3386880} $ & $  \frac{2431}{58060800} $ \\
	$ 2 $              & $0 $    & $  21 $  & $  -18 $          & $  \frac{77}{12} $  & $  -\frac{13}{10} $  & $  \frac{11}{64} $  & $  -\frac{2431}{151200} $ & $  \frac{2717}{2419200} $  & $  -\frac{221}{3628800} $  \\
	$ 3 $              & $0 $    & $0 $     & $  \frac{9}{2} $  & $  -\frac{11}{4} $  & $  \frac{117}{160} $ & $  -\frac{11}{96} $ & $  \frac{2431}{201600} $  & $  -\frac{247}{268800} $   & $  \frac{221}{4147200} $   \\
	$ 4 $              & $0 $    & $0 $     & $0 $              & $  \frac{11}{24} $  & $  -\frac{13}{60} $  & $  \frac{11}{240} $ & $  -\frac{221}{37800} $   & $  \frac{247}{483840} $    & $  -\frac{17}{518400} $    \\
	$ 5 $              & $0 $    & $0 $     & $0 $              & $0 $                & $  \frac{13}{480} $  & $  -\frac{1}{96} $  & $  \frac{221}{120960} $   & $  -\frac{19}{96768} $     & $  \frac{17}{1161216} $    \\
	$ 6 $              & $0 $    & $0 $     & $0 $              & $0 $                & $0 $                 & $  \frac{1}{960} $  & $  -\frac{17}{50400} $    & $  \frac{19}{376320} $     & $  -\frac{17}{3628800} $   \\
\hline
 \end{tabular}\label{tab:coeffG01s}
  \\
%
\begin{center}
 \begin{tabular} {c|ccccccccc}

	\multicolumn{9}{c}{\framebox{$ \gamma_{2}^{j,s} $}                                                                                          }                                                                                                                 \\
	$_j \backslash ^s$ & $  2 $              & $  4 $                & $  6 $               & $  8 $                  & $  10 $                  & $  12 $                    & $  14 $                        & $  16 $                          \\ \hline
	$ 1 $              & $  \frac{825}{4} $  & $  -\frac{2753}{16} $ & $  \frac{6313}{96} $ & $  -\frac{9303}{640} $  & $  \frac{20153}{9600} $  & $  -\frac{68959}{322560} $ & $  \frac{2192333}{135475200} $ & $  -\frac{9231079}{9754214400} $ \\
	$ 2 $              & $  -\frac{249}{2} $ & $  \frac{305}{2} $    & $  -\frac{421}{6} $  & $  \frac{139}{8} $      & $  -\frac{5207}{1920} $  & $  \frac{176803}{604800} $ & $  -\frac{55913}{2419200} $    & $  \frac{426439}{304819200} $    \\
	$ 3 $              & $0 $                & $  -\frac{709}{16} $  & $  \frac{1063}{32} $ & $  -\frac{2685}{256} $  & $  \frac{21953}{11520} $ & $  -\frac{52613}{230400} $ & $  \frac{14001}{716800} $      & $  -\frac{125099}{99532800} $    \\
	$ 4 $              & $0 $                & $0 $                  & $  -\frac{301}{48} $ & $  \frac{817}{240} $    & $  -\frac{2941}{3600} $  & $  \frac{101}{864} $       & $  -\frac{54847}{4838400} $    & $  \frac{4987}{6220800} $        \\
	$ 5 $              & $0 $                & $0 $                  & $0 $                 & $  -\frac{1817}{3840} $ & $  \frac{2323}{11520} $  & $  -\frac{37753}{967680} $ & $  \frac{17803}{3870720} $     & $  -\frac{72977}{195084288} $    \\
	$ 6 $              & $0 $                & $0 $                  & $0 $                 & $0 $                    & $  -\frac{637}{28800} $  & $  \frac{313}{40320} $     & $  -\frac{789}{627200} $       & $  \frac{38419}{304819200} $     \\
\hline
\end{tabular}

\begin{tabular} {c|ccccccccc}

	\multicolumn{9}{c}{\framebox{$ \gamma_{3}^{j,s} $}                                                                                        }                                          \\
	$_j \backslash ^s$ & $  1 $               & $  3 $               & $  5 $                 & $  7 $                  & $  9 $                   & $  11 $                       & $  13 $                        & $  15 $                        \\ \hline
	$ 1 $              & $  -\frac{1647}{4} $ & $  \frac{5561}{8} $  & $  -\frac{38129}{96} $ & $  \frac{11165}{96} $   & $  -\frac{40039}{1920} $ & $  \frac{2042161}{806400} $   & $  -\frac{4303871}{19353600} $ & $  \frac{6010433}{406425600} $ \\
	$ 2 $              & $  357 $             & $  -\frac{2993}{4} $ & $  \frac{23023}{48} $  & $  -\frac{36377}{240} $ & $  \frac{11029}{384} $   & $  -\frac{4400891}{1209600} $ & $  \frac{639353}{1935360} $    & $  -\frac{655343}{29030400} $  \\
	$ 3 $              & $0 $                 & $  \frac{2137}{8} $  & $  -\frac{8463}{32} $  & $  \frac{32879}{320} $  & $  -\frac{50911}{2304} $ & $  \frac{2464253}{806400} $   & $  -\frac{1911013}{6451200} $  & $  \frac{11453}{537600} $      \\
	$ 4 $              & $0 $                 & $0 $                 & $  \frac{5551}{96} $   & $  -\frac{3619}{96} $   & $  \frac{7553}{720} $    & $  -\frac{515197}{302400} $   & $  \frac{595907}{3225600} $    & $  -\frac{419393}{29030400} $  \\
	$ 5 $              & $0 $                 & $0 $                 & $0 $                   & $  \frac{1881}{320} $   & $  -\frac{6605}{2304} $  & $  \frac{300767}{483840} $    & $  -\frac{104549}{1290240} $   & $  \frac{2095}{290304} $       \\
	$ 6 $              & $0 $                 & $0 $                 & $0 $                   & $0 $                    & $  \frac{1991}{5760} $   & $  -\frac{54389}{403200} $    & $  \frac{77323}{3225600} $     & $  -\frac{19639}{7526400} $    \\
\hline
 \end{tabular}\label{tab:coeffG3js}\\
 
\vspace{10mm}
 \begin{tabular} {c|ccccccccc}

	\multicolumn{8}{c}{\framebox{$\gamma_{4}^{j,s} $}                                                                                          }                                                                                                          \\
	$_j \backslash ^s$ & $  2 $                 & $  4 $                  & $  6 $                    & $  8 $                     & $  10 $                         & $  12 $                        & $  14 $                            \\ \hline
	$ 1 $              & $  -\frac{47555}{32} $ & $  \frac{190709}{128} $ & $  -\frac{400233}{640} $  & $  \frac{1121033}{7680} $  & $  -\frac{23458073}{1075200} $  & $  \frac{35199901}{15482880} $ & $  -\frac{852863297}{4877107200} $ \\
	$ 2 $              & $  \frac{8687}{4} $    & $  -\frac{34773}{16} $  & $  \frac{4644}{5} $       & $  -\frac{425111}{1920} $  & $  \frac{1512829}{44800} $      & $  -\frac{1387199}{387072} $   & $  \frac{85408297}{304819200} $    \\
	$ 3 $              & $  -\frac{32127}{32} $ & $  \frac{185441}{128} $ & $  -\frac{934671}{1280} $ & $  \frac{8842181}{46080} $ & $  -\frac{100865411}{3225600} $ & $  \frac{9994067}{2867200} $   & $  -\frac{98657507}{348364800} $   \\
	$ 4 $              & $0 $                   & $  -\frac{23833}{64} $  & $  \frac{12277}{40} $     & $  -\frac{196591}{1920} $  & $  \frac{1944769}{100800} $     & $  -\frac{9191771}{3870720} $  & $  \frac{9035869}{43545600} $      \\
	$ 5 $              & $0 $                   & $0 $                    & $  -\frac{68941}{1280} $  & $  \frac{31835}{1024} $    & $  -\frac{998839}{129024} $     & $  \frac{17575681}{15482880} $ & $  -\frac{10930351}{97542144} $    \\
	$ 6 $              & $0 $                   & $0 $                    & $0 $                      & $  -\frac{23699}{5760} $   & $  \frac{737969}{403200} $      & $  -\frac{5213}{14336} $       & $  \frac{13298189}{304819200} $    \\
\hline
 \end{tabular}\label{tab:coeffG4js}\\
\vspace{10mm}
 \begin{tabular} {c|ccccccccc}

	\multicolumn{8}{c}{\framebox{$ \gamma_{5}^{j,s} $}   }                                                                                                                \\
	$_j \backslash ^s$ & $  1 $                & $  3 $                    & $  5 $                     & $  7 $                       & $  9 $                          & $  11 $                          & $  13 $                           \\ \hline
	$ 1 $              & $  \frac{22769}{16} $ & $  -\frac{1296283}{384} $ & $  \frac{4380193}{1920} $  & $  -\frac{42520013}{57600} $ & $  \frac{113941067}{806400} $   & $  -\frac{463361041}{25804800} $ & $  \frac{1135527679}{696729600} $ \\
	$ 2 $              & $  -\frac{13539}{4} $ & $  \frac{2484625}{384} $  & $  -\frac{1952053}{480} $  & $  \frac{58912711}{46080} $  & $  -\frac{292763719}{1209600} $ & $  \frac{2374678163}{77414400} $ & $  -\frac{53957761}{19353600} $   \\
	$ 3 $              & $  \frac{71825}{32} $ & $  -\frac{697647}{128} $  & $  \frac{9779869}{2560} $  & $  -\frac{14726891}{11520} $ & $  \frac{813614173}{3225600} $  & $  -\frac{850321511}{25804800} $ & $  \frac{854790599}{278691840} $  \\
	$ 4 $              & $0 $                  & $  \frac{81433}{48} $     & $  -\frac{3590417}{1920} $ & $  \frac{89371417}{115200} $ & $  -\frac{210499391}{1209600} $ & $  \frac{958855703}{38707200} $  & $  -\frac{854451569}{348364800} $ \\
	$ 5 $              & $0 $                  & $0 $                      & $  \frac{191637}{512} $    & $  -\frac{303739}{1152} $    & $  \frac{149339629}{1935360} $  & $  -\frac{200654617}{15482880} $ & $  \frac{400813631}{278691840} $  \\
	$ 6 $              & $0 $                  & $0 $                      & $0 $                       & $  \frac{8889011}{230400} $  & $  -\frac{8043643}{403200} $    & $  \frac{116210893}{25804800} $  & $  -\frac{105142393}{174182400} $ \\
\hline
\end{tabular}\label{tab:coeffG5js}
 \end{center}
 
\begin{tabular} {c|ccccccccc}

	\multicolumn{7}{c}{\framebox{$ \gamma_{6}^{j,s} $}                                                                                          }                                                                                                \\
	$_j \backslash ^s$ & $  2 $                    & $  4 $                        & $  6 $                       & $  8 $                           & $  10 $                            & $  12 $                              \\ \hline
	$ 1 $              & $  \frac{6360697}{1536} $ & $  -\frac{83914747}{15360} $  & $  \frac{205972691}{76800} $ & $  -\frac{238333495}{344064} $   & $  \frac{34401989173}{309657600} $ & $  -\frac{67943553227}{5573836800} $ \\
	$ 2 $              & $  -\frac{4527427}{384} $ & $  \frac{11995117}{960} $     & $  -\frac{85032277}{15360} $ & $  \frac{218343289}{161280} $    & $  -\frac{1086174427}{5160960} $   & $  \frac{7897450723}{348364800} $    \\
	$ 3 $              & $  \frac{6894791}{512} $  & $  -\frac{149105731}{10240} $ & $  \frac{67081587}{10240} $  & $  -\frac{2785868387}{1720320} $ & $  \frac{2911274687}{11468800} $   & $  -\frac{25580115979}{928972800} $  \\
	$ 4 $              & $  -\frac{1024813}{192} $ & $  \frac{818159}{96} $        & $  -\frac{22057967}{4800} $  & $  \frac{1273775}{1008} $        & $  -\frac{4112465311}{19353600} $  & $  \frac{421956389}{17418240} $      \\
	$ 5 $              & $0 $                      & $  -\frac{20343329}{10240} $  & $  \frac{54424859}{30720} $  & $  -\frac{644732521}{1032192} $  & $  \frac{7580219749}{61931520} $   & $  -\frac{8635714091}{557383680} $   \\
	$ 6 $              & $0 $                      & $0 $                          & $  -\frac{7421471}{25600} $  & $  \frac{1913875}{10752} $       & $  -\frac{133004803}{2867200} $    & $  \frac{816994861}{116121600} $     \\
\end{tabular}\label{tab:coeffG6js}
	\multicolumn{7}{c}{\framebox{$ \gamma_{7}^{j,s} $}                                                                                          }                                                                                                        \\
	$_j \backslash ^s$ & $  1 $                    & $  3 $                       & $  5 $                          & $  7 $                            & $  9 $                               & $  11 $                                 \\ \hline
	$ 1 $              & $  -\frac{1663559}{768} $ & $  \frac{121709581}{15360} $ & $  -\frac{312427823}{46080} $   & $  \frac{19673089399}{7741440} $  & $  -\frac{291425548999}{541900800} $ & $  \frac{2867948486117}{39016857600} $  \\
	$ 2 $              & $  \frac{4244105}{384} $  & $  -\frac{195851269}{7680} $ & $  \frac{811588697}{46080} $    & $  -\frac{12636131167}{2150400} $ & $  \frac{1402888481}{1209600} $      & $  -\frac{986035976887}{6502809600} $   \\
	$ 3 $              & $  -\frac{9859807}{512} $ & $  \frac{24914363}{640} $    & $  -\frac{4694903647}{184320} $ & $  \frac{11825097051}{1433600} $  & $  -\frac{23555895221}{14745600} $   & $  \frac{1149280143791}{5573836800} $   \\
	$ 4 $              & $  \frac{4053463}{384} $  & $  -\frac{3400213}{120} $    & $  \frac{1948423099}{92160} $   & $  -\frac{317943461}{43008} $     & $  \frac{11099163623}{7372800} $     & $  -\frac{280439920391}{1393459200} $   \\
	$ 5 $              & $0 $                      & $  \frac{121754083}{15360} $ & $  -\frac{1736132939}{184320} $ & $  \frac{31935948797}{7741440} $  & $  -\frac{19885290673}{20643840} $   & $  \frac{1102592661521}{7803371520} $   \\
	$ 6 $              & $0 $                      & $0 $                         & $  \frac{8062469}{4608} $       & $  -\frac{1689833129}{1290240} $  & $  \frac{4530651799}{11289600} $     & $  -\frac{1362209431217}{19508428800} $ \\
\hline
 \end{tabular}\label{tab:coeffG7js}\\
%
\begin{center}
 \begin{tabular} {c|ccccccccc}

	\multicolumn{6}{c}{\framebox{$ \gamma_{8}^{j,s} $}                                                                                          }                                                                                \\
	$_j \backslash ^s$ & $  2 $                         & $  4 $                         & $  6 $                              & $  8 $                                & $  10 $                                 \\ \hline
	$ 1 $              & $  -\frac{36667771}{6144} $    & $  \frac{137576377}{12288} $   & $  -\frac{58268412071}{8601600} $   & $  \frac{3483804392591}{1734082560} $ & $  -\frac{3962944769797}{11147673600} $ \\
	$ 2 $              & $  \frac{7932001}{256} $       & $  -\frac{3629434021}{92160} $ & $  \frac{62558288021}{3225600} $    & $  -\frac{393958246799}{77414400} $   & $  \frac{1159836156341}{1393459200} $   \\
	$ 3 $              & $  -\frac{1378786789}{20480} $ & $  \frac{5365102241}{73728} $  & $  -\frac{847780261361}{25804800} $ & $  \frac{3370548827407}{412876800} $  & $  -\frac{5974574495}{4644864} $        \\
	$ 4 $              & $  \frac{124022959}{1920} $    & $  -\frac{425217367}{5760} $   & $  \frac{13920972659}{403200} $     & $  -\frac{34003821749}{3870720} $     & $  \frac{122575184351}{87091200} $      \\
	$ 5 $              & $  -\frac{1391241377}{61440} $ & $  \frac{4788563821}{122880} $ & $  -\frac{114133290857}{5160960} $  & $  \frac{1565177035693}{247726080} $  & $  -\frac{609862623793}{557383680} $    \\
	$ 6 $              & $0 $                           & $  -\frac{154509701}{18432} $  & $  \frac{3653368319}{460800} $      & $  -\frac{105576999331}{36126720} $   & $  \frac{275386936571}{464486400} $     \\
\hline
\end{tabular}\label{tab:coeffG8js}

\begin{tabular} {c|ccccccccc}

	\multicolumn{6}{c}{\framebox{$ \gamma_{9}^{j,s} $}                                                                                          }                                                                                     \\
	$_j \backslash ^s$ & $  1 $                       & $  3 $                           & $  5 $                               & $  7 $                                  & $  9 $                                    \\ \hline
	$ 1 $              & $  \frac{72574433}{40960} $  & $  -\frac{33250133}{3072} $      & $  \frac{321566834083}{25804800} $   & $  -\frac{48892934193757}{8670412800} $ & $  \frac{424304933869777}{312134860800} $ \\
	$ 2 $              & $  -\frac{139238303}{7680} $ & $  \frac{2812070945}{49152} $    & $  -\frac{1815292535809}{38707200} $ & $  \frac{21601089607747}{1238630400} $  & $  -\frac{144839901437071}{39016857600} $ \\
	$ 3 $              & $  \frac{326578731}{5120} $  & $  -\frac{3318429817}{23040} $   & $  \frac{1712182537183}{17203200} $  & $  -\frac{13787028701293}{412876800} $  & $  \frac{295876625683021}{44590694400} $  \\
	$ 4 $              & $  -\frac{45278785}{512} $   & $  \frac{8612926109}{46080} $    & $  -\frac{4880466678209}{38707200} $ & $  \frac{17258868186769}{412876800} $   & $  -\frac{22984018946099}{2786918400} $   \\
	$ 5 $              & $  \frac{851962171}{20480} $ & $  -\frac{44593891643}{368640} $ & $  \frac{293291537465}{3096576} $    & $  -\frac{8497199910629}{247726080} $   & $  \frac{89611617850873}{12485394432} $   \\
	$ 6 $              & $0 $                         & $  \frac{713014643}{23040} $     & $  -\frac{41944969211}{1075200} $    & $  \frac{51535195929577}{2890137600} $  & $  -\frac{167924593843033}{39016857600} $ \\
\hline
 \end{tabular}\label{tab:coeffG9js}\\
%

 \begin{tabular} {c|ccccccccc}

	\multicolumn{5}{c}{\framebox{$ \gamma_{10}^{j,s} $}                                                                                         }                                                             \\
	$_j \backslash ^s$ & $  2 $                             & $  4 $                                 & $  6 $                                  & $  8 $                                       \\ \hline
	$ 1 $              & $  \frac{4188185133}{819200} $     & $  -\frac{668034376553}{45875200} $    & $  \frac{298635437603}{26214400} $      & $  -\frac{5008680379354717}{1248539443200} $ \\
	$ 2 $              & $  -\frac{69622578569}{1474560} $  & $  \frac{288361412803}{3686400} $      & $  -\frac{3733878109439}{82575360} $    & $  \frac{14728473880769}{1114767360} $       \\
	$ 3 $              & $  \frac{105043267843}{589824} $   & $  -\frac{90447675495017}{412876800} $ & $  \frac{176246672641693}{1651507200} $ & $  -\frac{2489696844203003}{89181388800} $   \\
	$ 4 $              & $  -\frac{72784945777}{230400} $   & $  \frac{560166907039}{1612800} $      & $  -\frac{49295772110719}{309657600} $  & $  \frac{13961643516553}{348364800} $        \\
	$ 5 $              & $  \frac{776330530877}{2949120} $  & $  -\frac{1732963120055}{5505024} $    & $  \frac{30099792470603}{198180864} $   & $  -\frac{705923122737949}{17836277760} $    \\
	$ 6 $              & $  -\frac{613186054801}{7372800} $ & $  \frac{436748965809}{2867200} $      & $  -\frac{37201323168383}{412876800} $  & $  \frac{1036259700960391}{39016857600} $    \\
\end{tabular}\label{tab:coeffG10js}
\vspace{10mm}
\begin{tabular} {c|ccccccccc}

	\multicolumn{5}{c}{\framebox{$ \gamma_{11}^{j,s} $}                                                                                         }                                                            \\
	$_j \backslash ^s$ & $  1 $                             & $  3 $                                 & $  5 $                                   & $  7 $                                     \\ \hline
	$ 1 $              & $  -\frac{4390861889}{4915200} $   & $  \frac{1961183621879}{206438400} $   & $  -\frac{38437694104309}{2477260800} $  & $  \frac{5499256218016573}{624269721600} $ \\
	$ 2 $              & $  \frac{12960937279}{737280} $    & $  -\frac{25335942630283}{309657600} $ & $  \frac{1196416018969}{14155776} $      & $  -\frac{1266083597711297}{34681651200} $ \\
	$ 3 $              & $  -\frac{20979287141}{184320} $   & $  \frac{22404670502027}{68812800} $   & $  -\frac{141336992746169}{550502400} $  & $  \frac{38716231419983}{412876800} $      \\
	$ 4 $              & $  \frac{561041955133}{1843200} $  & $  -\frac{6585350448109}{9676800} $    & $  \frac{2333996611139251}{4954521600} $ & $  -\frac{1416438955564847}{8918138880} $  \\
	$ 5 $              & $  -\frac{518643543809}{1474560} $ & $  \frac{2994059366393}{3870720} $     & $  -\frac{17684742866503}{33030144} $    & $  \frac{3766355622175117}{20808990720} $  \\
	$ 6 $              & $  \frac{16753583267}{115200} $    & $  -\frac{1461464779987}{3225600} $    & $  \frac{76285236828841}{206438400} $    & $  -\frac{2867374053933953}{20808990720} $ \\
\hline
\end{tabular}\label{tab:coeffG11js}
\begin{tabular} {c|ccccccccc}

	\multicolumn{4}{c}{\framebox{$ \gamma_{12}^{j,s} $}                                                                                         }                            \\
	$_j \backslash ^s$ & $  2 $                                   & $  4 $                                     & $  6 $                                      \\ \hline
	$ 1 $              & $  -\frac{112405393703}{39321600} $      & $  \frac{256194365128103}{19818086400} $   & $  -\frac{2434459781946581}{178362777600} $ \\
	$ 2 $              & $  \frac{14313033811823}{309657600} $    & $  -\frac{12498570084137}{117964800} $     & $  \frac{1677509997505583}{22295347200} $   \\
	$ 3 $              & $  -\frac{239571154081783}{825753600} $  & $  \frac{2904792531811861}{6606028800} $   & $  -\frac{120767420385113}{495452160} $     \\
	$ 4 $              & $  \frac{16651792420267}{19353600} $     & $  -\frac{3664938831889}{3538944} $        & $  \frac{3986190927755}{7962624} $          \\
	$ 5 $              & $  -\frac{91659779628347}{70778880} $    & $  \frac{1912786165948037}{1321205760} $   & $  -\frac{5996938416572659}{8918138880} $   \\
	$ 6 $              & $  \frac{32945086429187}{34406400} $     & $  -\frac{28119643773793}{23592960} $      & $  \frac{1463376996779051}{2477260800} $    \\
	$ 7 $              & $  -\frac{684201699957701}{2477260800} $ & $  \frac{10590857503323379}{19818086400} $ & $  -\frac{23381239709702027}{71345111040} $ \\
	$ 8 $              & $0 $                                     & $  -\frac{62410695774613}{619315200} $     & $  \frac{166206910739587}{1592524800} $     \\
	$ 9 $              & $0 $                                     & $0 $                                       & $  -\frac{577143489956477}{39636172800} $   \\
	\hline\\
%
%
	\multicolumn{4}{c}{\framebox{$ \gamma_{13}^{j,s} $}                                    }                                                                                \\
	$_j \backslash ^s$ & $  1 $                                 & $  3 $                                     & $  5 $                                       \\ \hline
	$ 1 $              & $  \frac{4480458061}{14745600} $       & $  -\frac{113509427927533}{19818086400} $  & $  \frac{4932648254178133}{356725555200} $   \\
	$ 2 $              & $  -\frac{692621429069}{61931520} $    & $  \frac{1586490110623811}{19818086400} $  & $  -\frac{811950375727963}{7431782400} $     \\
	$ 3 $              & $  \frac{104673184486151}{825753600} $ & $  -\frac{3256104685646191}{6606028800} $  & $  \frac{4461263791004999}{9512681472} $     \\
	$ 4 $              & $  -\frac{91340140787233}{154828800} $ & $  \frac{156492482258165}{99090432} $      & $  -\frac{14394426401891911}{11890851840} $  \\
	$ 5 $              & $  \frac{314766782631703}{247726080} $ & $  -\frac{2809659549121933}{990904320} $   & $  \frac{281042178651130801}{142690222080} $ \\
	$ 6 $              & $  -\frac{1355991878713}{1075200} $    & $  \frac{39617412502949}{13762560} $       & $  -\frac{336875717415931}{165150720} $      \\
	$ 7 $              & $  \frac{72172041424697}{154828800} $  & $  -\frac{15307727148091081}{9909043200} $ & $  \frac{233096036944318531}{178362777600} $ \\
	$ 8 $              & $0 $                                   & $  \frac{105529641662017}{309657600} $     & $  -\frac{376559490960019}{796262400} $      \\
	$ 9 $              & $0 $                                   & $0 $                                       & $  \frac{26397815834028073}{356725555200} $  \\
	\hline
\end{tabular}\label{tab:coeffG13js}

\begin{tabular} {c|ccccccccc}

	\multicolumn{3}{c}{\framebox{$ \gamma_{14}^{j,s} $}                                }                                                \\
	$_j \backslash ^s$ & $  2 $                                       & $  4 $                                          \\ \hline
	$ 1 $              & $  \frac{206796563366483}{184968806400} $    & $  -\frac{82215574092860581}{9988315545600} $   \\
	$ 2 $              & $  -\frac{618579830111987}{19818086400} $    & $  \frac{64337807855429791}{624269721600} $     \\
	$ 3 $              & $  \frac{8402777041707077}{26424115200} $    & $  -\frac{447073478617395173}{713451110400} $   \\
	$ 4 $              & $  -\frac{7430296569379097}{4954521600} $    & $  \frac{17606260552429}{8257536} $             \\
	$ 5 $              & $  \frac{6461512425366487}{1761607680} $     & $  -\frac{4349564850250920329}{998831554560} $  \\
	$ 6 $              & $  -\frac{222276140779005643}{46242201600} $ & $  \frac{3412415816069604403}{624269721600} $   \\
	$ 7 $              & $  \frac{253818565838765371}{79272345600} $  & $  -\frac{1685837673188956681}{407686348800} $  \\
	$ 8 $              & $  -\frac{7380514090083647}{8670412800} $    & $  \frac{540458699263271861}{312134860800} $    \\
	$ 9 $              & $0 $                                         & $  -\frac{1232692494304457819}{3995326218240} $ \\
	\hline
\end{tabular}\label{tab:coeffG14js}
\begin{tabular} {c|ccccccccc}

	\multicolumn{3}{c}{\framebox{$ \gamma_{15}^{j,s} $}    }                                                                      \\
	$_j \backslash ^s$ & $  1 $                                     & $  3 $                                      \\ \hline
	$ 1 $              & $  -\frac{3469384763227}{46242201600} $    & $  \frac{1187663466854671}{475634073600} $  \\
	$ 2 $              & $  \frac{19888331524783}{3963617280} $     & $  -\frac{127712690841809}{2264924160} $    \\
	$ 3 $              & $  -\frac{2547496947290741}{26424115200} $ & $  \frac{31541032831484593}{59454259200} $  \\
	$ 4 $              & $  \frac{4883476194008213}{6606028800} $   & $  -\frac{18839933887964653}{7431782400} $  \\
	$ 5 $              & $  -\frac{469284844441745}{176160768} $    & $  \frac{648190780936041043}{95126814720} $ \\
	$ 6 $              & $  \frac{6936784270529417}{1445068800} $   & $  -\frac{79914423777412117}{7431782400} $  \\
	$ 7 $              & $  -\frac{109591789469563}{26214400} $     & $  \frac{783868077139349141}{79272345600} $ \\
	$ 8 $              & $  \frac{17967831915895}{12845056} $       & $  -\frac{18230346413898349}{3715891200} $  \\
	$ 9 $              & $0 $                                       & $  \frac{30182462841350683}{29727129600} $  \\
\hline\\
 %

	\multicolumn{2}{c}{\framebox{$ \gamma_{16}^{j,s} $}}                                  \\
	$_j \backslash ^s$ & $  2 $                                           \\ \hline
	$ 1 $              & $  \frac{131482629287941}{407686348800} $        \\
	$ 2 $              & $  \frac{7661258420564069}{499415777280} $       \\
	$ 3 $              & $  -\frac{22226193309071407}{89181388800} $      \\
	$ 4 $              & $  \frac{5778045333852317}{3185049600} $         \\
	$ 5 $              & $  -\frac{1363999477757941669}{199766310912} $   \\
	$ 6 $              & $  \frac{35510799156457152389}{2497078886400} $  \\
	$ 7 $              & $  -\frac{94340848292247889769}{5707608883200} $ \\
	$ 8 $              & $  \frac{12526197981986130907}{1248539443200} $  \\
	$ 9 $              & $  -\frac{19789563585947876009}{7990652436480} $ \\
\hline
\end{tabular}\label{tab:coeffG16js}
 \clearpage
\begin{center}
\begin{table}\caption{$\lambda_j^{2i-1}$}
%
\label{tab:coeffLijs}
 \end{table}
 \clearpage

\begin{table}\caption{$\hat \lambda_j^{2i-1}$}
	\begin{tabular} {c|ccccccc}
		\multicolumn{5}{c}{\framebox{$ \hat \lambda_j^{2i-1}$ }                                           }                                \\
		$_j \backslash ^{2i-1}$ & $ 1 $            & $ 3 $              & $ 5 $                 & $ 7 $                   \\ \hline
		$ 1 $              & $ -\frac{9}{8} $ & $ -\frac{45}{32} $ & $ -\frac{999}{512} $  & $ -\frac{10043}{4096} $ \\
		$ 2 $              & $0$              & $ -\frac{81}{16} $ & $ -\frac{45}{16} $    & $ -\frac{653}{128} $    \\
		$ 3 $              & $0$              & $0$                & $ -\frac{8427}{512} $ & $ \frac{4929}{4096} $   \\
		$ 4 $              & $0$              & $0$                & $0$                   & $ -\frac{5929}{128} $   \\
	\hline
	\end{tabular}\\
	\end{table}
	\vspace{10mm}
	\begin{table}
\begin{tabular}{c|ccccccc}

	\multicolumn{6}{c}{\framebox{ $ \hat \lambda_j^{2i-1}$ }   }                                                                                                                                                                             \\
	$_j \backslash ^{2i-1}$ & $ 9 $                        & $ 11 $                        & $ 13 $                                & $ 15 $                                  & $ 17 $                                        \\ \hline
	$ 1 $              & $ -\frac{968501}{327680} $   & $ -\frac{22665719}{6553600} $ & $ -\frac{3488799931}{880803840} $     & $ -\frac{2201362888069}{493250150400} $ & $ -\frac{705270944819797}{142056043315200} $  \\
	$ 2 $              & $ -\frac{1861}{320} $        & $ -\frac{140657}{20480} $     & $ -\frac{50772389}{6451200} $         & $ -\frac{458120711}{51609600} $         & $ -\frac{1338763001}{135475200} $             \\
	$ 3 $              & $ -\frac{8061723}{655360} $  & $ -\frac{24220341}{2621440} $ & $ -\frac{2503082907}{209715200} $     & $ -\frac{310605553233}{23488102400} $   & $ -\frac{5548949079117}{375809638400} $       \\
	$ 4 $              & $ \frac{154}{5} $            & $ -\frac{510211}{12800} $     & $ -\frac{112849}{23040} $             & $ -\frac{534407009}{25804800} $         & $ -\frac{244133119}{12902400} $               \\
	$ 5 $              & $ -\frac{15717645}{131072} $ & $ \frac{77607165}{524288} $   & $ -\frac{82475458145}{528482304} $    & $ \frac{443312655515}{8455716864} $     & $ -\frac{16937466865475}{315680096256} $      \\
	$ 6 $              & $0$                          & $ -\frac{30089667}{102400} $  & $ \frac{75371433}{143360} $           & $ -\frac{24124153401}{40140800} $       & $ \frac{7345300413}{20070400} $               \\
\hline
\end{tabular}\label{tab:coeffLhatijs}
 \end{table}
\end{center}	 
\begin{table} \caption{$\eta_i^{j,s}$}
\begin{tabular} {c|ccccccccc}
	\multicolumn{10}{c}{\framebox{$ \eta_{-1}^{j,s} $}   }                                                                                                                                                                                                   \\
	$_j \backslash ^s$ & $  2 $            & $  4 $            & $  6 $            & $  8 $             & $  10 $              & $  12 $                & $  14 $                  & $  16 $                   & $  18 $                       \\ \hline
	$ 1 $              & $  -\frac{3}{2} $ & $  \frac{3}{2} $  & $  -\frac{5}{8} $ & $  \frac{7}{48} $  & $  -\frac{7}{320} $  & $  \frac{11}{4800} $   & $  -\frac{143}{806400} $ & $  \frac{143}{13547520} $ & $  -\frac{2431}{4877107200} $ \\
	$ 2 $              & $0 $              & $  -\frac{3}{4} $ & $  \frac{1}{2} $  & $  -\frac{7}{48} $ & $  \frac{1}{40} $    & $  -\frac{11}{3840} $  & $  \frac{143}{604800} $  & $  -\frac{143}{9676800} $ & $  \frac{221}{304819200} $    \\
	$ 3 $              & $0 $              & $0 $              & $  -\frac{1}{8} $ & $  \frac{1}{16} $  & $  -\frac{9}{640} $  & $  \frac{11}{5760} $   & $  -\frac{143}{806400} $ & $  \frac{13}{1075200} $   & $  -\frac{221}{348364800} $   \\
	$ 4 $              & $0 $              & $0 $              & $0 $              & $  -\frac{1}{96} $ & $  \frac{1}{240} $   & $  -\frac{11}{14400} $ & $  \frac{13}{151200} $   & $  -\frac{13}{1935360} $  & $  \frac{17}{43545600} $      \\
	$ 5 $              & $0 $              & $0 $              & $0 $              & $0 $               & $  -\frac{1}{1920} $ & $  \frac{1}{5760} $    & $  -\frac{13}{483840} $  & $  \frac{1}{387072} $     & $  -\frac{17}{97542144} $     \\
	$ 6 $              & $0 $              & $0 $              & $0 $              & $0 $               & $0 $                 & $  -\frac{1}{57600} $  & $  \frac{1}{201600} $    & $  -\frac{1}{1505280} $   & $  \frac{17}{304819200} $     \\
\end{tabular}\label{tab:coeffPminus1js}
%

	\multicolumn{10}{c}{\framebox{$ \eta_{0}^{j,s} $}            }                                                                                                                                                              \\
	$_j \backslash ^s$ & $  1 $ & $  3 $   & $  5 $            & $  7 $            & $  9 $             & $  11 $              & $  13 $                 & $  15 $                  & $  17 $                     \\ \hline
	$ 1 $              & $  6 $ & $  -12 $ & $  \frac{15}{2} $ & $  -\frac{7}{3} $ & $  \frac{7}{16} $  & $  -\frac{11}{200} $ & $  \frac{143}{28800} $  & $  -\frac{143}{423360} $ & $  \frac{2431}{135475200} $ \\
	$ 2 $              & $0 $   & $  6 $   & $  -6 $           & $  \frac{7}{3} $  & $  -\frac{1}{2} $  & $  \frac{11}{160} $  & $  -\frac{143}{21600} $ & $  \frac{143}{302400} $  & $  -\frac{221}{8467200} $   \\
	$ 3 $              & $0 $   & $0 $     & $  \frac{3}{2} $  & $  -1 $           & $  \frac{9}{32} $  & $  -\frac{11}{240} $ & $  \frac{143}{28800} $  & $  -\frac{13}{33600} $   & $  \frac{221}{9676800} $    \\
	$ 4 $              & $0 $   & $0 $     & $0 $              & $  \frac{1}{6} $  & $  -\frac{1}{12} $ & $  \frac{11}{600} $  & $  -\frac{13}{5400} $   & $  \frac{13}{60480} $    & $  -\frac{17}{1209600} $    \\
	$ 5 $              & $0 $   & $0 $     & $0 $              & $0 $              & $  \frac{1}{96} $  & $  -\frac{1}{240} $  & $  \frac{13}{17280} $   & $  -\frac{1}{12096} $    & $  \frac{17}{2709504} $     \\
	$ 6 $              & $0 $   & $0 $     & $0 $              & $0 $              & $0 $               & $  \frac{1}{2400} $  & $  -\frac{1}{7200} $    & $  \frac{1}{47040} $     & $  -\frac{17}{8467200} $    \\
\end{tabular}\label{tab:coeffP0js}
 \end{table}
%
\clearpage
 \begin{tabular} {c|ccccccccc}

	\multicolumn{9}{c}{\framebox{$ \eta_{1}^{j,s} $}                               }                                                                                                                                                                         \\
	$_j \backslash ^s$ & $  2 $             & $  4 $                & $  6 $                & $  8 $                   & $  10 $                & $  12 $                     & $  14 $                      & $  16 $                         \\ \hline
	$ 1 $              & $  \frac{345}{8} $ & $  -\frac{1401}{32} $ & $  \frac{3661}{192} $ & $  -\frac{17717}{3840} $ & $  \frac{911}{1280} $  & $  -\frac{49159}{645120} $  & $  \frac{60203}{10035200} $  & $  -\frac{784927}{2167603200} $ \\
	$ 2 $              & $  -\frac{63}{2} $ & $  42 $               & $  -\frac{1015}{48} $ & $  \frac{169}{30} $      & $  -\frac{149}{160} $  & $  \frac{3971}{37800} $     & $  -\frac{83369}{9676800} $  & $  \frac{13}{24192} $           \\
	$ 3 $              & $0 $               & $  -\frac{429}{32} $  & $  \frac{675}{64} $   & $  -\frac{8977}{2560} $  & $  \frac{5123}{7680} $ & $  -\frac{89353}{1075200} $ & $  \frac{94757}{12902400} $  & $  -\frac{15041}{30965760} $    \\
	$ 4 $              & $0 $               & $0 $                  & $  -\frac{101}{48} $  & $  \frac{283}{240} $     & $  -\frac{281}{960} $  & $  \frac{1307}{30240} $     & $  -\frac{5213}{1209600} $   & $  \frac{1511}{4838400} $       \\
	$ 5 $              & $0 $               & $0 $                  & $0 $                  & $  -\frac{87}{512} $     & $  \frac{569}{7680} $  & $  -\frac{28417}{1935360} $ & $  \frac{13711}{7741440} $   & $  -\frac{911}{6193152} $       \\
	$ 6 $              & $0 $               & $0 $                  & $0 $                  & $0 $                     & $  -\frac{1}{120} $    & $  \frac{1}{336} $          & $  -\frac{11089}{22579200} $ & $  \frac{53}{1058400} $         \\
\end{tabular}\label{tab:coeffP1js}
	\multicolumn{9}{c}{\framebox{$ \eta_{2}^{j,s} $}      }                                                                                                                                                                                                   \\
	$_j \backslash ^s$ & $  1 $              & $  3 $              & $  5 $                & $  7 $                 & $  9 $                    & $  11 $                       & $  13 $                      & $  15 $                        \\ \hline
	$ 1 $              & $  -\frac{327}{4} $ & $  \frac{1233}{8} $ & $  -\frac{9565}{96} $ & $  \frac{5159}{160} $  & $  -\frac{12007}{1920} $  & $  \frac{651761}{806400} $    & $  -\frac{68783}{921600} $   & $  \frac{1261403}{243855360} $ \\
	$ 2 $              & $  \frac{357}{4} $  & $  -\frac{789}{4} $ & $  \frac{401}{3} $    & $  -\frac{3587}{80} $  & $  \frac{34483}{3840} $   & $  -\frac{1444399}{1209600} $ & $  \frac{2431}{21504} $      & $  -\frac{99619}{12441600} $   \\
	$ 3 $              & $0 $                & $  \frac{641}{8} $  & $  -\frac{2611}{32} $ & $  \frac{10467}{320} $ & $  -\frac{83851}{11520} $ & $  \frac{839597}{806400} $    & $  -\frac{224003}{2150400} $ & $  \frac{111917}{14515200} $   \\
	$ 4 $              & $0 $                & $0 $                & $  \frac{925}{48} $   & $  -\frac{6143}{480} $ & $  \frac{10469}{2880} $   & $  -\frac{182537}{302400} $   & $  \frac{53989}{806400} $    & $  -\frac{18643}{3483648} $    \\
	$ 5 $              & $0 $                & $0 $                & $0 $                  & $  \frac{403}{192} $   & $  -\frac{11953}{11520} $ & $  \frac{110479}{483840} $    & $  -\frac{39017}{1290240} $  & $  \frac{149}{54432} $         \\
	$ 6 $              & $0 $                & $0 $                & $0 $                  & $0 $                   & $  \frac{1493}{11520} $   & $  -\frac{20609}{403200} $    & $  \frac{4939}{537600} $     & $  -\frac{41137}{40642560} $   \\
\hline
 \end{tabular}%
\label{tab:coeffP16js}
 
 \end{center}
 
\end{document}